\newif\ifarXiv

\documentclass[review]{elsarticle}
\usepackage{lineno,hyperref}
\modulolinenumbers[5]

\journal{Elsevier}

\usepackage{here}
\usepackage{amsmath,amssymb,amsfonts,bm}
\usepackage[ruled, vlined, linesnumbered]{algorithm2e}
\usepackage[noend]{algorithmic}
\usepackage{graphicx}
\usepackage{textcomp}
\usepackage{xcolor}
\usepackage{xspace}

\usepackage{lscape}

\usepackage{comment}
\usepackage{url}
\usepackage{subcaption}

\newcommand{\memo}[1]{}
\newcommand{\bmC}{\bm{C}}
\newcommand{\bmM}{\bm{M}}
\newcommand{\bmX}{\bm{X}}
\newcommand{\bmS}{\bm{S}}
\newcommand{\bmU}{\bm{U}}
\newcommand{\bmV}{\bm{V}}
\newcommand{\bmH}{\bm{H}}

\newcommand{\bmP}{\bm{P}}
\newcommand{\bmtheta}{\bm{\theta}}
\newcommand{\bmD}{\bm{D}}

\newcommand{\trans}{^{\top}}

\newcommand{\inmathbbr}[2]{\in \mathbb{R}^{#1 \times #2}}

\newcommand{\argmin}{\mathop{\rm argmin}\limits}

\newcommand{\mem}{node-class membership}
\newcommand{\Mem}{Node-class membership}
\newcommand{\attcor}{attribute-class correlation}
\newcommand{\Attcor}{Attribute-class correlation}

\newcommand{\seiji}[1]{#1}
\newcommand{\seijirv}[1]{#1}

\newcommand{\nonl}{\renewcommand{\nl}{\let\nl\oldnl}}

\newtheorem{definition}{Definition}[section]

\begin{document}

\begin{frontmatter}

\title{
GenCAT: Generating Attributed Graphs with Controlled Relationships between Classes, \\ Attributes, and Topology
}
\author[OU]{Seiji Maekawa\corref{mycorrespondingauthor}}
\author[OU]{Yuya Sasaki}
\author[TUe]{George Fletcher}
\author[OU]{Makoto Onizuka}

\cortext[mycorrespondingauthor]{Corresponding author, maekawa.seiji@ist.osaka-u.ac.jp}

\address[OU]{Osaka University, 1-5 Yamadaoka,Suita, Osaka, Japan}
\address[TUe]{Eindhoven University of Technology, P.O. Box 513, MB, Eindhoven 5600,The Netherlands}

\begin{abstract}
Generating large synthetic attributed graphs with node labels is an important task to support various experimental studies for graph analytic methods. 
Existing graph generators fail to simultaneously simulate \seijirv{core/border and homophily/heterophily phenomena which real-world graphs exhibit, i.e., the relationships between labels, attributes, and topology}.
Motivated by this limitation, we propose GenCAT, an attributed graph generator for controlling those relationships, which has the following advantages.
    (i) GenCAT generates graphs with user-specified node degrees and flexibly controls the relationship between nodes and labels by incorporating the connection proportion for each node to classes. 
    (ii) Generated attribute values follow user-specified distributions, and users can flexibly control the correlation between the attributes and labels. 
    (iii) Graph generation scales linearly to the number of edges.
GenCAT is the first generator to support all three of these practical features\seijirv{, i.e., it can capture both core/border and homophily/heterophily phenomena while ensuring its scalability}.
Through extensive experiments, we demonstrate that GenCAT can efficiently generate high-quality complex attributed graphs with user-controlled relationships between labels, attributes, and topology.

\end{abstract}


\end{frontmatter}

\section{Introduction} \label{sec:introduction}
Graph is a fundamental data structure consisting of nodes and edges. Graphs are ubiquitous in many application domains, such as web graph \cite{flake2002self}, social networks \cite{fortunato2010community}, 
computer vision~\cite{jain2016structural}, 
and gene expressions~\cite{ben1999clustering,kulis2009semi}.
Nodes of real-world graphs often have attributes (e.g., blogs in web graphs have text information).
Of the rich variety of attributed graph analytic methods, graph clustering~\cite{bojchevski2018bayesian,ye2018deep,ijcai2019-509}, classification~\cite{bhagat2011node,krishna2020factorized}, and subgraph matching~\cite{agarwal2020chisel} are examples of techniques widely used in the graph mining and management fields~\cite{Sahu2017ubiquity}. 
Researchers and developers often validate the effectiveness and efficiency of new graph analytic methods by using multiple graphs with various characteristics 
in order to clarify their applicability and/or limitations.

Although many real-world attributed graphs are available in repositories such as SNAP~\cite{snapnets}, they typically lack ground-truth node labels (i.e., an assignment of nodes to groups)
and are often small scale. 
Hence, most available real-world graphs are often not applicable to empirically evaluate graph mining and management methods. 
In this paper, we call a set of nodes with the same label a \textbf{\textit{class}} and assume that nodes in a class tend to share similar attributes. 

Consequently, it is important to use large synthetic attributed graphs with class labels\footnote{\seiji{We focus on undirected and unweighted graphs in this paper since many existing graph analytic methods are designed for such graphs.}}  which exhibit various characteristics of real-world graphs for studying graph mining and management methods. 
The major topological characteristics that web graph and social networks have are small-world property and power law node degree distribution. 
Regarding the interplay between classes, attributes, and topology, we focus on two widely known phenomena in attributed graphs: \textbf{\textit{core/border}} and \textbf{\textit{network homophily/heterophily}}. 

\subsection{Phenomena in attributed graphs}
\noindent{\bf Core/border.}
There are two kinds of nodes in a class, core and border~\cite{ester1996dbscan}.
The core nodes in a class are nodes with similar attribute values to the average attribute values of the nodes in the class.
The border nodes in a class are nodes with attribute values mixed from nodes in different classes. 
In other words, core and border nodes strongly and weakly belong to their classes, respectively. 
To generalize these phenomena, we assume that \seiji{each node is correlated with multiple classes with certain degrees}\footnote{A class label represents the class to which a node is most correlated. \seiji{That is, we do not consider overlapping or hierarchical classes. An expansion to graphs with overlapping or hierarchical classes is our future work as we will discuss in Section \ref{sec:conclusion}.}}. 

\begin{figure}[!t]
  \centering
  \includegraphics[trim={0mm 0mm 0mm 0},width=6.cm]{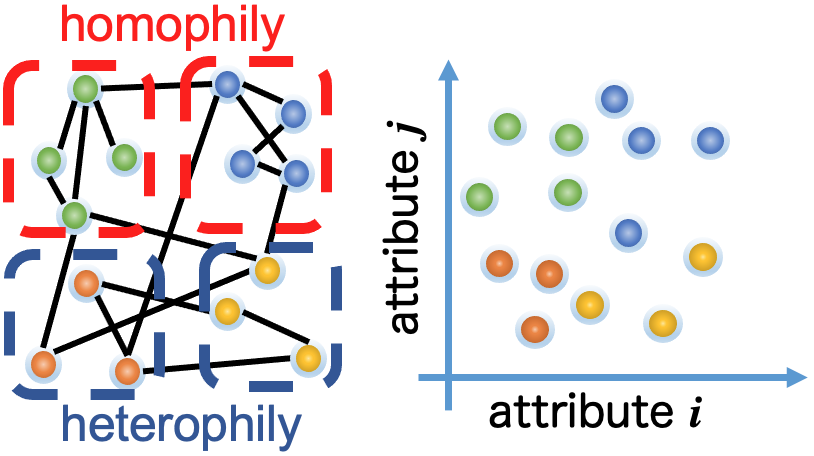}
  \caption{
  Colors indicate the classes which nodes belong to. 
  The nodes in each class have similar attributes. 
  The nodes in a class with the homophily property (surrounded by a red circle) tend to connect internally.
  The nodes in a class with the heterophily property (surrounded by a blue circle) tend to connect externally.
}
  \label{fg: topology type}
\end{figure}
\smallskip
\noindent{\bf Homophily/heterophily.}
Homophily and heterophily are phenomena to express the relationships between classes and the topology. 
First, homophily is a well-known phenomenon of real-world graphs where nodes in the same class are more likely to connect to each other. 
Classes with the homophily property are called communities, i.e., sets of nodes that are densely connected internally \cite{mcpherson2001birds}.
Second, heterophily is the inverse notion of homophily: nodes with dissimilar attributes are more likely to connect to each other. 
From the viewpoint of classes, nodes in a class with heterophily property are more likely to connect to other classes than the class. 
Figure~\ref{fg: topology type} shows an intuitive example of classes with homophily/heterophily properties. 
Recent graph analytic methods~\cite{zhu2020graph,zhu2020beyond} aim to capture both homophily/heterophily phenomena. 

In summary, to capture the relationships between classes, attributes, and topology, graph generators should support the core/border and homophily/\!\! heterophily phenomena. 
We call these relationships \textbf{\textit{class structure}}.

\subsection{Existing generators}
From the above discussion, graph generators should support class structure. 
However, most existing techniques do not explicitly generate class labels and control the class structure of graphs. 
\seijirv{In particular, traditional graph generators \cite{erdHos1960evolution,albert2002statistical,aiello2000random,seshadhri2012community} focus on statistics for a whole graph, e.g., node degree distributions, and not on the class structure. As a result, they cannot control connection proportions between classes. }
gMark~\cite{BBCFLA16gmark,bagan2017gmark}, pgMark~\cite{pgmark2018}, and TrillionG~\cite{park2017trilliong} support class labels but are not designed to control the class structure. 
LFR~\cite{lancichinetti2008benchmark} and DC-SBM~\cite{karrer2011stochastic} are designed to support the relationship between the class labels and topology for evaluating the performance of community detection~\cite{ye2018danmf} and label propagation methods~\cite{krishna2020factorized}. 
They cannot simulate the core/border phenomena because they do not control the connection proportions to classes from individual nodes (i.e., all nodes in a class have the same connection proportions).
Some other methods for generating attributed graphs with class labels have been proposed, such as ANC~\cite{largeron2015generating} and DANCer~\cite{benyahia2016dancer}. 
These attributed graph generators cannot control the homophily/heterophily phenomena in each class since they can set the same ratio of intra- and inter-edges to all classes. 
Moreover, ANC and DANCer cannot control the variety of connection proportions between classes but each pair of classes in real-world graphs may have different connection proportions.

Graph generators based on deep learning have also been proposed recently, such as VGAE~\cite{kipf2016variational}, GraphVAE~\cite{simonovsky2018graphvae}, and NetGAN~\cite{bojchevski2018netgan}. 
They aim to reproduce synthetic graphs from given input graphs, but they do not support generating graphs with user-controlled class structure.

In summary, there are no graph generators that support flexibly controlling the proportions of connections between nodes and classes. 

\subsection{Technical Challenges}
Given the practical need for generating attributed graphs with class labels and the shortcomings of the state of the art,
two major difficulties must be overcome to realize a new generator for supporting core/border and homophily/heterophily phenomena: ensuring 1) the high-quality of generated graphs and 2) the efficiency of graph generation, so as to capture the characteristics of real-world attributed graphs and generate graphs with user-controlled class structure in a reasonable time. 
    First, to generate high-quality graphs, we need to satisfy constraints simultaneously on various attributed graph characteristics, such as node degree distributions, attribute value distributions, and the class structure which exhibits core/border and homophily/heterophily phenomena. 
    Second, graph generators should scale well to generate massive graphs for wide applicability.
Moreover, these two aspects, the high-quality and efficiency of graph generation, are in a trade-off relationship. 
It is not trivial to satisfy the constraints of the graph generation problem since they are interdependent and the problem of satisfying even the single constraint on node degrees is NP-complete~\cite{bagan2017gmark}.
Hence, the generator should be designed to balance the two aspects.

\subsection{Contributions}
\label{ssec: contribution}
We propose a new graph generator, GenCAT, for synthetic graph generation to solve these challenges. 
GenCAT allows us to flexibly control the class structure of a generated attributed graph by capturing core/border and homophily/heterophily phenomena. 
We introduce \textbf{\textit{node-class membership proportion}} and \textbf{\textit{node-class connection proportion}} to capture the core/border and the homophily/heterophily phenomena, respectively. 
First, the node-class membership proportion of a node represents how likely the node belongs to classes. 
A core node has a value close to 1 for its class and a border node has balanced values for multiple classes.
Second, the node-class connection proportion of a node represents how likely the node connects to classes. 
The node-class membership and connection proportions of a node should be similar if the node's class has the homophily property and be opposite if the node's class has the heterophily property. 

GenCAT efficiently generates the high-quality graphs by utilizing latent factors which are intermediate data structures to capture the class structure of attributed graphs. 
To improve the quality of generated graphs, GenCAT generates edges that follow the user-specified class structure by solving the constraints expressed with latent factors for node-class membership/connection proportions.
GenCAT also generates attributes that follow the user-specified class structure by solving constraints expressed with latent factors capturing the correlation between attributes and classes.
To efficiently generate graphs, we heuristically assign priorities to the constraints of graph characteristics to be satisfied and take an effective edge generation approach that utilizes inverse transform sampling~\cite{devroye2006nonuniform}. 
Thanks to this approach, GenCAT can generate graphs in linear time to the number of edges.

We summarize the properties of graph generators in Table~\ref{tb: property}. 
GenCAT supports all of the desired properties, whereas existing methods lack one or more of them. 
Since GenCAT is a general generator that can support various settings, it can also simulate the existing generators LFR and DC-SBM by giving appropriate parameters (the details are described in Section~\ref{ssec: simulating}) without any modifications of GenCAT itself.

The three main characteristics of GenCAT are as follows:
\vspace{-0.5mm}
\begin{itemize}
    \item GenCAT generates graphs with user-specified node degrees and the connection proportions between nodes and classes. 
    \vspace{-0.5mm}
    \item The attribute values generated by GenCAT follow user-specified distributions and users can flexibly control the correlation between the attributes and classes. 
    \vspace{-0.5mm}
    \item GenCAT scales linearly to the number of generated edges and can generate graphs with billion edges.
\end{itemize}
\vspace{-0.5mm}
GenCAT is the first method having all three of these desirable characteristics\seijirv{, i.e., it can generate realistic graphs exhibiting both core/border and homophily/heterophily phenomena while ensuring its scalability}.
For community use and further study, our complete code base is available as open source.\footnote{\url{https://github.com/seijimaekawa/GenCAT}}

\subsection{Organization}
The rest of this paper is organized as follows. 
We describe the problem statement and the challenges of a graph generation problem in Section~\ref{sec: problem statement}. 
We propose GenCAT in Section~\ref{sec:proposal}. 
Section \ref{sec:experiments} gives a detailed experimental analysis of the quality of generated graphs and the efficiency of graph generation.  
We also position GenCAT with respect to the state of the art in Section \ref{sec:related}. 
We give concluding remarks in Section~\ref{sec:conclusion}.

\section{Problem statement}
\label{sec: problem statement}
In this section, we first describe notation and define the graph/class features of attributed graphs. Then, we define our problem and describe challenges to solve our problem. 

\subsection{Notation}
An {\em undirected attributed graph with class labels} is expressed with a triple $G=(\bmS,\bmX,\bmC)$ where $\bmS \in \{0,1\}^{n\times n}$ is an adjacency matrix, $\bmX\in \mathbb{R}^{n\times d}$ is an attribute matrix assigning attributes to nodes, ${\bmC} \in \{1,...,k\}^{n}$ is a vector assigning each node to a single class label\footnote{As we mention in Section~\ref{sec:introduction}, we assume that each node is correlated with multiple classes with some degree so a class label indicates the class to which each node is most correlated. }, and $n,d,k$ are the numbers of nodes, attributes, and classes, respectively. If there is an edge between nodes $i$ and $j$, $\bmS_{ij}$ and $\bmS_{ji}$ are set to one. 
We define $\bmS_{i.}$ and $\bmS_{.j}$ as the $i$-th row and the $j$-th column of $\bmS$, respectively.
Also, we define $\Omega_l$ as the class for label $l$ (i.e., the set of nodes labeled with $l$). 

\subsection{Graph features}
We highlight two features that real-world graphs typically have:
topology statistics and attribute statistics, which are desirable in synthetically generated graphs.

\begin{landscape}
\begin{table}[t]
  \caption{Properties of graph generators: $\checkmark$ and $\times$ indicate that the generators satisfy and do not satisfy the characteristic, respectively. 
  The topological reproducibility (the second right-most column) indicates whether generators can reproduce input graphs.
  The complexity column indicates the time complexity including training steps.
  $O(m$+$)$ represents the complexity which is $O(m)$ when the number of nodes is fixed. *: the time complexity of ANC and DANCer also depends on the cost of a clustering method, e.g., the cost of KMedoids is $O(n^2k)$. $p$ represents the number of threads, $f$ the dimension of the latent factors and $B$ the batch size. 
  See Table  \ref{tb: input} for other parameters.}
  \label{tb: property}
\setlength{\tabcolsep}{3pt}
  \begin{center}
	\begin{tabular}{l|ccccc} 
      Generator &  \begin{tabular}{c} Node-class membership\\ proportion \end{tabular} & Class labels (class size) & \begin{tabular}{c} Node\\ attribute\end{tabular} & \begin{tabular}{c} Topological\\ reproducibility\end{tabular} & Complexity \\ \hline
        LFR \cite{lancichinetti2008benchmark}& $\checkmark$(class-level only) &  $\checkmark$(power law only) & $\times$ & $\times$ & $O(m$+$)$ \\ 
        DC-SBM~\cite{karrer2011stochastic} &$\checkmark$(class-level only) & $\checkmark$(input list) & $\times$ & $\times$ & $O(n^2)$ \\ 
        ANC~\cite{largeron2015generating}& $\checkmark$(class-level only) &$\checkmark$(clustering method-driven) & $\checkmark$ & $\times$ & $O(m)^*$\\ 
        DANCer~\cite{benyahia2016dancer}& $\checkmark$(class-level only) &$\checkmark$(clustering method-driven) & $\checkmark$ & $\times$ & $O(m)^*$ \\ 
        pgMark~\cite{BBCFLA16gmark,bagan2017gmark,pgmark2018} & $\times$ & $\checkmark$(schema-driven) & $\checkmark$ & $\times$ & $O(m)$ \\ 
        TrillionG~\cite{park2017trilliong} & $\times$ & $\checkmark$(schema-driven) & $\times$ & $\times$ & $O((m\log n) / p)$ \\ 
        VGAE~\cite{kipf2016variational} & $\times$ & $\times$ & $\times$ & $\checkmark$ & $O(fn^2)$ \\ 
        GraphVAE~\cite{simonovsky2018graphvae} & $\times$ & $\times$ & $\times$ & $\checkmark$ & $O(n^4)$ \\ 
        NetGAN~\cite{bojchevski2018netgan} & $\times$ & $\times$ & $\times$ & $\checkmark$ & $O(B)$ \\ 
        GenCAT & $\checkmark$ & $\checkmark$(power law, normal, and input list) & $\checkmark$ & $\checkmark$ &
        $O(mkr+dkn)$\\ 
 	\end{tabular}
 \end{center}
\end{table}
\end{landscape}

\noindent{\bf Topology statistics.}
Real graphs have well-known topological statistics \cite{leskovec2010kronecker,newman2010intro}:
for example, node degrees in social graphs often follow a power law distribution. 
For this reason, graph generators should support various types of distributions of node degrees. 

\vspace{0.2cm}
\noindent{\bf Attribute statistics.}
Attributes in real datasets typically follow underlying distributions. 
For example, binary categories follow Bernoulli distribution, such as word appearance and questions with two possible answers. 
Also, it is well known that many numerical attributes follow a normal distribution, such as biological data and data including measurement errors. 
So, graph generators should support various types of the distribution of attribute values such as Bernoulli and normal distributions, which support typical attributes. 

\subsection{Class features}
\label{ssec: basic statistics}
As we mentioned in Section~\ref{sec:introduction}, graph generators should support the node-class membership/connection proportions capturing the phenomena of core/border and homophily/heterophily. 
To generate the latent factors for expressing the node-class membership/connection proportions, 
we identify three basic statistics of class features as input parameters: class preference mean, class preference deviation, and class size distribution. 
Also, to generate the latent factor for expressing the relationship between the attributes and classes, we identify a statistic of class features for the attributes: \attcor.

\smallskip
\noindent{\bf Class preference mean.}
To simulate the homophily/heterophily phenomena, we introduce \textbf{\textit{class preference mean}}, $\bmM\inmathbbr{k}{k}$. 
An element of class preference mean, $\bmM_{l_1l_2}$ expresses the average of connection proportions from the nodes in class $l_1$ to the nodes in class $l_2$. 
We formulate class preference mean between class $l_1$ and class $l_2$ as follows:
\begin{align}
    \bmM_{l_1l_2} = \frac{1}{|\Omega_{l_1}|}\sum_{i\in \Omega_{l_1}}(\sum_{j\in \Omega_{l_2}}\bmS_{ij}/\sum_{j=1}^{n} \bmS_{ij}).
\end{align}
Class preference mean is a more general notion than the simple binary representation of homophily/heterophily. 
For example, if $\bmM_{ll} = 0.7$ and $k=3$, class $l$ has a stronger homophily property because the nodes in class $l$ are more likely to connect to each other than the nodes in other classes. 
Note that the diagonal elements express the proportions of the connections inside of each class. 

\smallskip
\noindent{\bf Class preference deviation.}
We also introduce \textbf{\textit{class preference deviation}}, $\bmD\inmathbbr{k}{k}$, in order to simulate the core/border phenomena. 
Class preference deviation indicates the variety of \mem\ proportions between classes. 
That is, it expresses the extent to which nodes in a class belong to multiple classes. 
Class preference deviation is a more general notion than the simple binary representation of core/border. 
An element of class preference deviation, $\bmD_{l_1l_2}$ indicates the standard deviation of connection proportions from nodes in class $l_1$ to nodes in class $l_2$.
We formulate class preference deviation between class $l_1$ and class $l_2$ as follows:
\begin{align}
    \bmD_{l_1l_2} = \sqrt{\frac{1}{|\Omega_{l_1}|}\sum_{i\in\Omega_{l_1}}(\sum_{j\in \Omega_{l_2}}\bmS_{ij}/\sum_{j=1}^{n} \bmS_{ij}-\bmM_{l_1l_2})^2}.
\end{align}
That is, class preference mean and class preference deviation express the average and deviation of the connection proportions between classes, respectively. 

\smallskip
\noindent{\bf Class size distribution.}
The class preference mean and class preference deviation capture detailed characteristics of classes, however, they lack information of class sizes. 
We introduce \textbf{\textit{class size distribution}}, which complements them. 
In many real-world graphs, such as social networks, the distribution of class sizes is usually well-approximated by power law \cite{palla2005uncovering,clauset2004finding}.

\smallskip
\noindent \textbf{\Attcor.}
In this paper, nodes in the same class tend to share similar attribute values as we mentioned in Section~\ref{sec:introduction}. 
Since the relationship between the attributes and classes is typically various, we assume that each attribute is correlated to classes with certain degrees. 
To capture the correlation, we introduce \textbf{\textit{\attcor}}, $\bmH\inmathbbr{d}{k}$, which represents the strength of the correlation between the attributes and classes.
We formulate \attcor\ between attribute $\delta$ and class $l$ as the average of values of attribute $\delta$ of nodes in class $l$: 
\begin{align}
    \bmH_{\delta l} = \frac{1}{|\Omega_l|}\sum_{i\in\Omega_l}(\bmX_{i\delta}).
\end{align}
Since we assume that the topology and the attributes share the same class structure, we utilize \mem\ proportions and the \attcor\ in order to capture the correlations between nodes and attributes.

\subsection{Problem definition and Challenges}
\label{ssec: challenges}

We can now define the problem that we solve in this paper. 
We assume two practical usage scenarios as follows. 
In the first scenario, the user inputs statistics of graphs to be generated so as to flexibly control the characteristics of generated graphs.
In the second scenario, the user inputs graphs with class labels so as to generate graphs similar to the given input graphs.

\smallskip
\noindent
{\bf Problem Definition.}
We give two definitions for these two usage scenarios. 
Given either 1) statistics of graphs; graph features (node degree and attribute distributions) and class features (class preference mean, class preference deviation, class size distribution, and \attcor)
\footnote{To make the input easier, we can accept the diagonal elements of class preference mean instead of class preference mean and class preference deviation. In this case, the deviation is randomly generated since we generate class preference mean from dirichlet distribution. }
or 2) topological information (node degree, class preference mean, class preference deviation and class size distribution) extracted from adjacency matrix $\bmS'$ and class labels $\bmC'$ of a given graph, attribute distributions, and \attcor, we efficiently generate $G=(\bmS,\bmX,\bmC)$ whose statistics are similar to the inputs.

To address this graph generation problem, we define the loss $L$ that indicates the difference between the user-specified statistics and the statistics of a generated graph. 
We formulate the loss as follows:
\begin{align}
\label{eq: overall loss}
    L & = L^\text{topo} + L^\text{attr}
\end{align}
where $L^\text{topo}$ indicates the loss of the topological part and $L^\text{attr}$ indicates the loss of the attribute part. 
We design the losses, $L^\text{topo}$ and $L^\text{attr}$, in order to clarify the relationship between the graph generation problem and the constraints that generated graphs should satisfy graph features and class features. 
First, we formulate the loss $L^\text{topo}$ as follows: 
\begin{align}
\label{eq: loss topo}
    L^\text{topo} = L^\text{topo}_{\text{graph feature}} + L^\text{topo}_{\text{class feature}}
\end{align}
where $L^\text{topo}_{\text{graph feature}}$ indicates the topological loss between the given graph features and the graph features in the generated graph and $L^\text{topo}_{\text{class feature}}$ indicates a topological loss between the given class features and the class features in the generated graph.
Second, we formulate the loss $L^\text{attr}$ as follows: 
\begin{align}
\label{eq: loss attr}
    L^\text{attr} = L^\text{attr}_{\text{graph feature}} + L^\text{attr}_{\text{class feature}}
\end{align}
where $L^\text{attr}_{\text{class feature}}$ indicates the attribute loss between the given graph features and the graph features in the generated graph and $L^\text{attr}_{\text{class feature}}$ indicates an attribute loss between the given class features and the class features in the generated graph. 

\smallskip
\noindent
{\bf Challenges.} 
To solve our problem, we address two major challenges. 
The first challenge is that GenCAT must generate edges that satisfy multiple topological constraints\footnote{The attribute part has fewer constraints such as the distribution of attribute values, so we focus on the topology part here.} of the graph features and the class features.

By adopting latent factors that express the node-class membership/connection proportions, $L^\text{topo}_{\text{class feature}}$ is expressed with two parts: 1) the loss between a generated graph and latent factors and 2) the loss between latent factors and class features. 
We formulate $L^\text{topo}_{\text{class feature}}$ as follows:
\begin{align}
\label{eq: loss class feature}
    L^\text{topo}_{\text{class feature}} = 
    \underbrace{L_{\text{edge precision}}}_{\text{between generated graph and latent factors}}   \nonumber \\ 
    + \underbrace{L_
    \text{mean} + L_\text{deviation} + L_\text{class size}}_{\text{between latent factors and class features}}
\end{align}
where 
$L_{\text{edge precision}}$ is a loss that expresses the precision of the generated edges according to the probabilities of the edge existence calculated by the latent factors, 
$L_\text{mean}$ is a loss between user-specified class preference mean and the class preference mean estimated from the latent factors, 
$L_\text{deviation}$ is a loss between user-specified class preference deviation and the class preference deviation estimated from the latent factors, 
and
$L_\text{class size}$ is a loss between the expected class size distribution and the class size distribution in the generated graph.

The second challenge is efficiently generating large scale graphs. 
There are two problems; 1) deciding whether there exists a graph satisfying given node degrees is an NP-complete problem~\cite{bagan2017gmark}, 
2) since GenCAT assumes that each node has its node-class membership/connection proportion, it is required to estimate the probabilities of the edge existence of all node pairs, resulting in the large cost of $O(kn^2)$.
To overcome the first problem, we design an efficient algorithm that heuristically assigns degrees to nodes and generates edges that satisfy the node degrees so as to avoid the large cost of satisfaction problem of graph generation. 
As for the second problem, we take an efficient edge generation approach that utilizes an approximation in order to achieve a linear time algorithm to the number of edges.

\section{GenCAT: attributed graph generator for controlling class structure}
\label{sec:proposal}
In this section, we explain the design of GenCAT. 
First of all, we introduce latent factors and design the loss for class features $L^\text{topo}_\text{class feature}$ consisting of $L_\text{edge precision}$, $L_\text{mean}$, $L_\text{deviation}$, and $L_\text{class size}$, shown in Eq.~\eqref{eq: loss class feature}  by using the latent factors. 
Also, we design the loss with graph feature, $L^\text{topo}_{\text{graph feature}}$ and provide an overview of graph generation by GenCAT. 
Then, we design the losses with class and graph features regarding attributes so that an adjacency matrix and an attribute matrix share the user-controlled class structure in Section~\ref{ssec: generating model}. 
As we described in Section \ref{sec: problem statement}, 
we provide two scenarios of graph generation. 
In the first scenario, users input graph features (node degree distribution and attribute value distribution) and class features (class preference mean, class preference deviation, class size distribution, and \attcor). 
In the second scenario, GenCAT extracts the features from an input graph.
We explain the first scenario in Section~\ref{ssec: algorithm} and then the second scenario in Section~\ref{ssec: reproduction}. 
Next, we analyze the time and space complexities of GenCAT in Section~\ref{ssec: complexity}. 
Finally, we show how GenCAT simulates existing generators in Section~\ref{ssec: simulating}. 
Table~\ref{tb: variables} lists the main symbols and their definitions for the following descriptions.
\begin{table}[t]
  \caption{Definition of main symbols.}
  \vspace{-3mm}
  \label{tb: variables}
\setlength{\tabcolsep}{5pt}
  \begin{center}
	\begin{tabular}{l|l}
	Variable & Explanation \\ \hline
    $\bmS \in\{0,1\}^{n\times n}$ &  adjacency matrix \\ 
    $\bmX \inmathbbr{n}{d}$ &  attribute matrix \\
    $\bmC \in \{1,\dots,k\}^{n}$ & class label \\ 
    $\Omega_l \in \{1,\dots,n\}^*$ & set of nodes labeled with class $l$ \\
    $\bmU \inmathbbr{n}{k}$ &  \mem\ proportions \\ 
    $\bmU' \inmathbbr{n}{k}$ & node-class connection proportions \\ 
    $\bmV \inmathbbr{d}{k}$ &  attribute-class proportions \\ 
    ${\bm \theta},{\bm \theta'} \in \mathbb{R}^{n}$ & expected and actual node degree proportions \\ 
    ${\bm \rho} \in \mathbb{R}^{k}_+$ & class size distribution \\ 
    $Z \subset \{1,\dots,k\}^{*}$ & candidate set for edge generation \\ 
 	\end{tabular}
 \end{center}
\end{table}
GenCAT supports various input parameters as shown in Table~\ref{tb: input}.
\begin{table}[t]
  \caption{Description of the graph generator parameters. }
  \vspace{-3mm}
  \label{tb: input}
\setlength{\tabcolsep}{5pt}
  \begin{center}
	\begin{tabular}{l|l} 
        Input &  Description \\ \hline
        $n,m,d,k \in \mathbb{N}$\hspace{-1mm} & number of nodes, edges, attributes, classes \\ 
        $\bmM \inmathbbr{k}{k}$ & class preference mean \\ 
        $\bmD \inmathbbr{k}{k}$ & class preference deviation \\ 
        $\bmH \inmathbbr{d}{k}$ & \attcor \\ 
        $\phi_C \in \mathbb{R}_+$ & parameter for class size distribution \\ 
        $\omega \in \mathbb{R}$ & deviation of normal distribution for attributes\\ 
        $r \in \mathbb{N}$ & number of iterations for edge generation \\ 
 	\end{tabular}
 \end{center}
\end{table}
The most basic parameters are 1) the number of nodes, edges, attributes, and classes to generate, and 2) the distribution parameters for class sizes\footnote{
As for the exponents, we choose typical values of real networks: $1\le \phi_C \le 2$, where $\phi_C$ is the parameter for class size \cite{lancichinetti2008benchmark}.}
and attribute values. 

\begin{figure}[t]
  \centering
  \includegraphics[trim={0mm 0mm 0mm 0},width=12.cm]{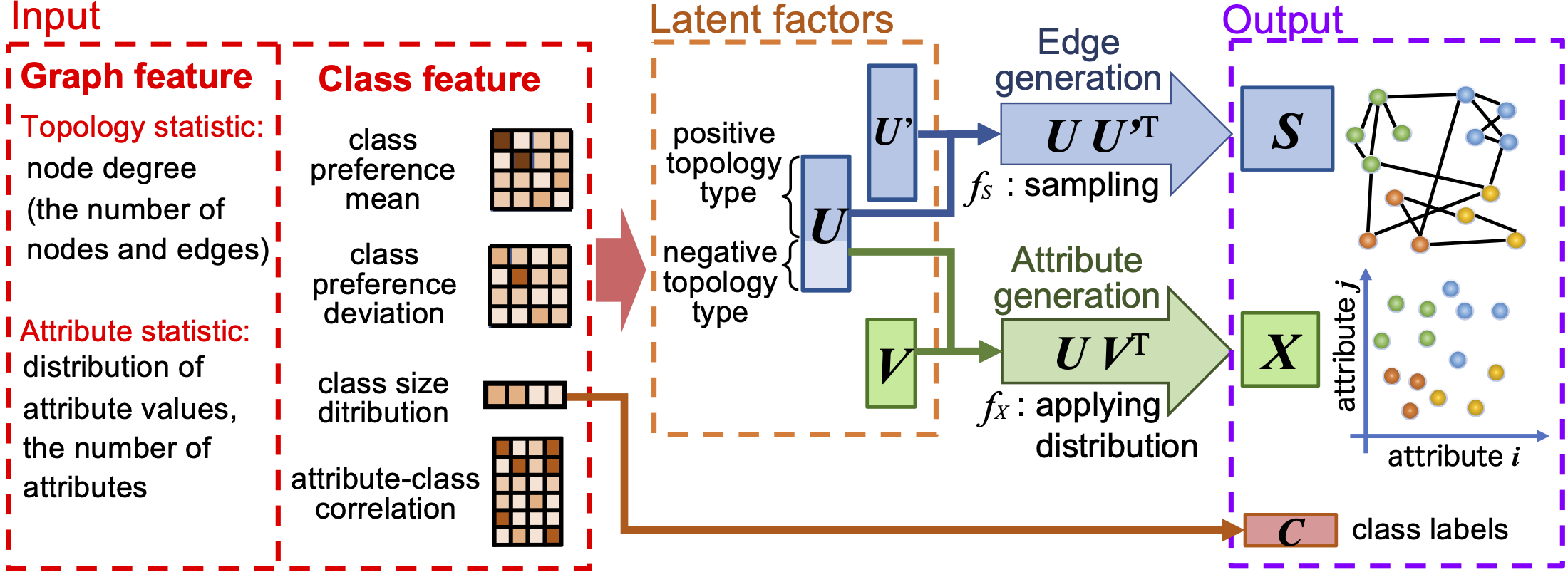}
  \caption{Illustration of our method. 
  GenCAT first generates intermediate data structures $\bmU,\bmU',\bmV$ and then generates the adjacency matrix $\bmS$ and the attribute matrix $\bmX$ by using $\bmU,\bmU',\bmV$. The class labels $\bmC$ are generated based on class size distribution.  }
  \label{fg: model_fg}
\end{figure}
\subsection{Generating Model}
\label{ssec: generating model}
The basic idea of GenCAT is to capture class structure by using intermediate data structures, called \textit{latent factors}, and then to generate graphs from the latent factors (See Figure \ref{fg: model_fg}).  

\subsubsection{Latent factors} \label{sssec:latent_factors}
GenCAT generates output graphs with class labels by using three latent factors: \mem\ proportions $\bmU \inmathbbr{n}{k}$, node-class connection proportions $\bmU' \inmathbbr{n}{k}$, and attribute-class proportions $\bmV \inmathbbr{d}{k}$, that are core components for capturing class features in the real-world graphs. 

GenCAT calculates adjacency matrix $\bmS$ according to the probabilities of edge existence between nodes by multiplying $\bmU$ and ${\bmU'}\trans$ and attribute matrix $\bmX$ by multiplying $\bmU$ and ${\bmV}\trans$.
By sharing $\bmU$ in the generations of $\bmS$ and $\bmX$, GenCAT can inject the common characteristics of classes into both the adjacency and the attribute matrices. 
We describe the detailed designs of $\bmU$ and $\bmU'$ in Section~\ref{sssec:loss from class feature} and $\bmV$ in Section~\ref{sssec:loss_attribute}.

We formally define three latent factors as follows:
\begin{definition}[\Mem\ proportions $\bmU$]
$\bmU$ regards a \\ class assignment projection from nodes to classes.
$\bmU$ is a matrix whose size is $n \times k$ and each element is in [0,1]. 
An element at $i$-th row and $j$-th column indicates that node $i$ likely belongs to class $j$ if the element is close to one.
\end{definition}

\begin{definition}[Node-class connection proportions $\bmU'$]
$\bmU'$ regards an \\ edge connectivity projection from nodes to classes. 
$\bmU'$ is a matrix whose size is $n \times k$ and each element is [0,1]. 
An element at $i$-th row and $j$-th column indicates that node $i$ likely has edges with nodes in class $j$ if the element is close to one. 
\end{definition}

\begin{definition}[Attribute-class proportions $\bmV$]
$\bmV$ regards an attribute \\ projection from attributes to classes. 
$\bmV$ is a matrix whose size is $d \times k$ and each element is [0,1]. 
An element at $i$-th row and $j$-th column indicates that attribute $i$ is likely correlated with class $j$ if the element is close to one.
\end{definition}

\subsubsection{Loss from class feature regarding topology}
\label{sssec:loss from class feature}
To incorporate the class features into the edge generation of GenCAT, we design the loss, $L^\text{topo}_{\text{class feature}}$ (Eq.~\eqref{eq: loss class feature}), which is expressed with 1) the loss between generated graph and latent factors and 2) the loss between latent factors and class features as follows.

\smallskip
\noindent{\bf Between generated graph and latent factors.}
We design GenCAT to generate edges by using latent factors, $\bmU$ and $\bmU'$. 
According to the definition of $\bmU$ and $\bmU'$, we can calculate the edge probability between nodes by multiplying $\bmU$ and ${\bmU'}\trans$, which expresses a composition of two projections, class assignment from nodes to classes ($\bmU$) and edge connectivity from classes to nodes (${\bmU'}\trans$). 
By using the edge probabilities, we formulate the loss that expresses the precision of the generated edges as below:
\begin{align}
    L_\text{edge precision} = \|\bmS-\bmU{\bmU'}\trans\|_F.
    \label{eq: L edge}
\end{align}
Recall the definition of homophily/heterophily in Section \ref{sec:introduction}: homophily is a phenomenon where the nodes with similar attributes are more likely to connect to each other and heterophily is the inverse notion of homophily.
By following them, we can design $\bmU$ and $\bmU'$ to have the same proportions for nodes (rows) with homophily property and to have the reverse proportions for nodes with heterophily property, respectively. 
To give precise definitions of classes with the homophily/heterophily properties, we introduce types of class. 
We assign \textit{positive topology type} to a class with the homophily property, and \textit{negative topology type} to a class with the heterophily property from input parameters as follows.
Positive topology type is assigned to class $l$ if the diagonal elements of the class preference mean are larger than a random connection proportion ($\bmM_{ll} \ge 1/k$), otherwise negative topology type is assigned. 
From the above discussion, we formulate $\bmU'$ of which nodes (rows) in positive topology type classes have the same values of $\bmU$ and nodes in negative topology type classes have the reversed values of $\bmU$ as below: 
\begin{eqnarray}
    \label{eq: inverseU}
    \bmU'_{i.}=\left\{
\begin{array}{ll}
 \bmU_{i.} & (i \in \Omega_{l} \text{\ and\ } \bmM_{ll} \ge 1/k)\\
 f_{\rm rev}(\bmU_{i.}) & (i \in \Omega_{l} \text{\ and\ } \bmM_{ll} < 1/k).
\end{array}
\right.
\end{eqnarray}
Also, the reverse function $f_{\rm rev}$ is formulated as follows: 
\begin{eqnarray}
    \label{eq: U prime}
    f_{\rm rev}(\bmU_{i.})=\left\{
\begin{array}{ll}
 1 - \bmU_{ih} & (h = l)\\
 (1-\bmU_{ih})\frac{\bmU_{il}}{\sum^k_{j \ne l}(1-\bmU_{ij})} & (h \ne l)
\end{array}
\right.
\end{eqnarray}
where node $i$ is a node whose class is typed with negative topology and class $l$ is a class which node $i$ belongs to. 
Also, by regularizing the values other than in the class each node belongs to, the sum of each row of $\bmU'$ equals to one. 

Notice that the edge generation by $\bmU{\bmU'}\trans$ is a more generalized form of $\bmU\bmU\trans$ used by SymNMF~\cite{kuang2012symmetric,kuang2015symnmf}, a well known technique for graph clustering. SymNMF is a special case when all classes have homophily property, so node-class membership proportions $\bmU$ are identical with node-class connection proportions $\bmU'$.

\smallskip
\noindent{\bf Between latent factors and class features.}
Next, we design loss formulas between latent factors and class features so that the latent factors precisely capture the class features, which are class preference mean, class preference deviation, and class size distribution. 
First, to reduce the loss between user-specified class preference mean and the class preference mean estimated by latent factors which is calculated by $\bmU_{i.}*\bmU'_{i.}$ as an approximation, we formulate the loss as follows:
\begin{align}
 \label{eq: loss M}
  L_\text{mean} = \sum_{l}^k\|\bmM_{l.} - \underbrace{\frac{1}{|\Omega_l|} \sum_{i\in\Omega_l}\bmU_{i.}*\bmU'_{i.}}_{\text{estimated connection proportions}}\|_F
\end{align}
where $*$ denotes the element-wise product. 
The reason that we adopt the approximation to calculate the estimated class preference mean is that the cost to obtain the exact proportion of edges between classes is the large cost of $O(kn^2)$\footnote{The cost comes from $\bmU{\bmU'}\trans$ which calculates all possible combinations of nodes.}. 
Next, to reduce the loss between user-specified class preference deviation and the estimated deviation, 
we also formulate the loss as follows: 
\begin{align}
 \label{eq: loss D}
  L_\text{deviation} = \sum_{l}^k\|\bmD_{l.} - \underbrace{\sqrt{\frac{1}{|\Omega_{l}|}\sum_{i\in\Omega_l}(\bmU_{i.}*\bmU'_{i.} - \overline{\bmU_{i.}*\bmU'_{i.}})^2}}_{\text{estimated deviation}}\|_F
\end{align}
where $\overline{\bmU_{i.}*\bmU'_{i.}}$ denotes the average of $\bmU_{i.}*\bmU'_{i.}$ where $i\in\Omega_l$. 
We adopt an approximation due to the same reason as Eq.~\eqref{eq: loss M}. 
Finally, we formulate the loss between the expected class sizes and the class sizes in a generated graph. 
The loss, $L_\text{class size}$, is described as: 
\begin{align}
\label{eq: class size}
    L_\text{class size} = \sum^k_{l=1}({\bm \rho}[l] - |\Omega_l|/n)^2
\end{align}
where ${\bm \rho}$ denotes the class size distribution specified by users and $|\Omega_l|/n$ represents the class size proportion of class $l$ in a generated graph.

\subsubsection{Loss from graph feature regarding topology}
\label{sssec:loss from graph feature}
The graph feature is expressed with the topology statistic and the attribute statistic and we focus on the topology part here. 
Because graph generators should support user-specified node degrees, 
the quality of a generated graph is computed by the difference between user-specified node degrees and the node degrees of a generated graph as follows:
\begin{align}
    L^\text{topo}_{\text{graph feature}} = \frac{1}{n}\sum^{n}_{i=1}|\frac{\bmtheta_i-\bmtheta'_i}{\bmtheta_i}|
    \label{eq: L degree}
\end{align}
where ${\bm \theta}$ is the expected node degree and ${\bm \theta'}$ is the actual node degrees. 
We employ \textit{Mean Absolute Percentage Error} (MAPE) for the loss. 
The reason that we utilize MAPE is that it can treat high degree nodes and low degree nodes equally. 
Other measures can be used, such as mean squared error. 

\subsubsection{Edge generation}
\label{sssec: graph generation}
Adjacency matrix $\bmS$ is generated by $\bmU{\bmU'}\trans$ (See Eq.~\eqref{eq: L edge}). 
Thanks to the latent factors, $\bmU$ and $\bmU'$, GenCAT can take into account user-specified class features in the edge generation. 
As we mentioned in Section~\ref{ssec: challenges}, the adjacency matrix generation has two problems; 1) deciding whether there exists a graph having given node degrees is an NP-complete problem, 
2) the computational cost for the probability of edge existence, $\bmU{\bmU'}\trans$, is $O(kn^2)$. 
Moreover, these problems are correlated by the dependency between node degrees (Eq.~\eqref{eq: L degree}) and edges generated based on the node-class membership/connection proportions (Eq.~\eqref{eq: L edge}) because the degree of node $i$ represents the number of edges associated with node $i$. 
Hence, GenCAT adopts an efficient algorithm that heuristically assigns degrees to nodes and generates edges that satisfy the node degrees so as to avoid the large cost of the satisfaction problem of graph generation. 
Also, to overcome the large cost to compute the edge probabilities, we incorporate an inverse transform sampling~\cite{devroye2006nonuniform} into our heuristic approach. 

\smallskip
\noindent{{\bf Approach to generate edges.}}
The ideas of the approach are two-fold: 1) we generate edges in order starting from high degree nodes and 2) we accelerate the calculation of the probability of edge existence by utilizing an inverse transform sampling.
The purpose of the first idea is that we avoid leaving high degree nodes to the later phase of edge generation so that high degree nodes can have a sufficient number of candidates to connect. 
If high degree nodes are left to the later phase of generation, most edges which should be associated with nodes may not be generated. 
Hence, we start the edge generation in the order starting from high degree nodes so that they have many candidates to connect with at the earlier phase of the edge generation. 
As for the second idea, we interpret the probability of the edge existence between node $i$ and node $j$ as computed by $\bmU_{i.}*\bmU'_{j.}$ in two selection steps, \textit{Class selection from source node} step based on $\bmU$ and \textit{Target node selection from class} step based on ${\bmU'}\trans$\footnote{We use source and target even if we assume undirected graphs}. 
Thanks to the interpretation that transforms the calculation into the two probabilistic selection steps, we can incorporate inverse transform sampling into the approach. 
First, we describe the two selection steps in detail, and then we explain how to utilize the sampling method. 

In \textit{Class selection from source node} step, we choose classes $Z\subset\{1,\dots,k\}^{\bmtheta_i}$ from each node $i$'s \mem\ proportions $\bmU_{i.}$. 
Then, in \textit{Target node selection from class} step, a node is selected from the node-class connection proportions ${\bmU'_{.l}}\trans$ for each class $l\in Z$. 
An edge between the selected node and node $i$ is generated\footnote{The edge probability between node $i$ and node $j$ is calculated by $\sum_{l=1}^k\bmU_{il}\bmU_{jl}$ which corresponds to $(\bmU{\bmU'}\trans)_{ij}$. }. 
This approach still suffers from the high computational cost to execute \textit{Target node selection from class} for each source node, which requires $O(kn^2)$. 

\begin{figure}[t]
  \centering
  \includegraphics[trim={0mm 0mm 0mm 0},width=8.cm]{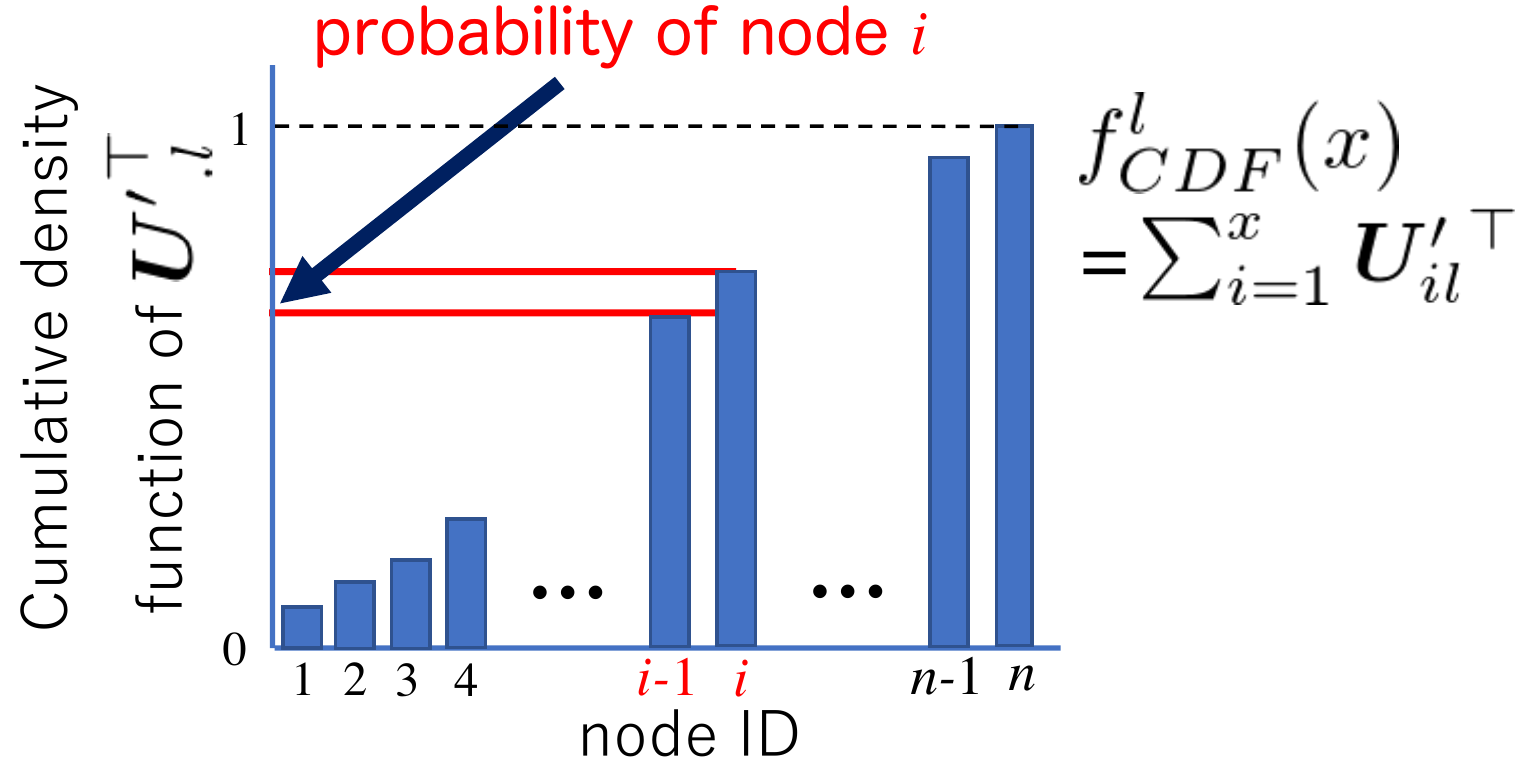}
  \caption{The concept of inverse transform sampling. Bars represent the cumulative density function of ${\bmU'}\trans_{.l}$ where $l$ indicates a class.} 
  \label{fg: CDF}
\end{figure}
\smallskip
\noindent{{\bf Inverse transform sampling.}}
To accelerate the approach, we incorporate an inverse transform sampling for monotone densities into the approach. 
It can considerably improve the speed of edge generation since it realizes that the time complexity of \textit{Target node selection from class} step is $O(1)$. 
First, we calculate the cumulative distribution function (CDF) of ${\bmU'}\trans$ for each class. 
Figure~\ref{fg: CDF} depicts the cumulative density function of $\bmU'$ by regularizing $\bmU'$ for a certain column. 
The horizontal axis indicates nodes, the vertical axis indicates the CDF of a column (class)  of $\bmU'$, and the function $f_{CDF}$ represents the CDF of the probability that nodes are selected from the class. 
The width between node $i-1$ and node $i$ indicates the probability of selecting node $i$, which is expressed with $f_{CDF}(i)-f_{CDF}(i-1)$. 
We generate the value of inverse CDF, $f_{CDF}^{-1}(x)$, which is a node ID of node $i$ if $f_{CDF}(i-1) \le x < f_{CDF}(i)$. 
This means that the inverse CDF returns a node ID based on the probability by choosing a random number from zero to one. 
For the quick access, we generate a list by which we can obtain a node ID based on its selecting probability, in a similar way as the inverse CDF, by dividing the range of random numbers into small steps. 
That is, a node ID of node $i$ is obtained by utilizing the list if the list receives the value between $f_{CDF}(i-1)$ and $f_{CDF}(i)$. 
Using this list, \textit{Target node selection from class} is executed in $O(1)$. 
Note that the complexity to generate the list is $O(kn)$ because we compute the CDF for each class. 

\subsubsection{Loss regarding attribute}
\label{sssec:loss_attribute}
As we mentioned in Section~\ref{sec: problem statement}, GenCAT should support both 1) the class feature that indicates the \attcor\ and 2) the graph feature for attributes, which is the distribution of attribute values. 

First, we incorporate the \mem\ proportions $\bmU$ into attribute generation so that GenCAT generates graphs with the class structure shared by the topology and the attributes. 
To share the class structure with the topology, an attribute matrix $\bmX$ is generated based on $\bmU\bmV\trans$, which expresses a composition of class assignment projection from nodes to classes ($\bmU$) and attribute projection from classes to attributes (${\bmV}\trans$). 
In order to reduce the loss between user-specified \attcor\ and the estimated \attcor, we formulate the loss as follows:
\begin{align}
\label{eq:loss_att_class_feature}
    L^\text{attr}_\text{class feature} =
    \sum_{l=1}^k\|\bmH_{.l}\trans-\underbrace{\frac{1}{|\Omega_l|}\sum_{i\in\Omega_l}(\bmU\bmV\trans)_{i.}}_\text{estimated \attcor}\|_F.
\end{align}
By designing attribute-class proportions $\bmV$ to reduce $L^\text{attr}_\text{class feature}$, generated attributes satisfy user-specified \attcor. 
Since \mem\ proportions are commonly used in edge generation and attribute generation, the generated attributes share the class structure with the topology.

Second, we design the loss $L^\text{attr}_\text{graph feature}$ so that the distributions of generated attribute values are similar to the user-specified distributions. 
We adopt the \textit{Earth-Mover} (EM) distance\footnote{Earth Mover distance is also known as Wasserstein-$1$} that is a widely used measurement of the distance between two distributions \cite{arjovsky2017wasserstein}. 
By using the EM distance, we formulate $L^\text{attr}_\text{graph feature}$ as follows:
\begin{align}
\label{eq:loss_att_feature}
    L^\text{attr}_{\text{graph feature}} = \sum_{\delta}^d \inf_{\gamma\in\prod(\bmX_{.\delta},p(\bmX_{.\delta}))} \mathbb{E}_{(x,y)\sim\gamma}[|x-y|]
\end{align}
where $\prod(\bmX_{i\delta},p(\bmX_{i\delta}))$ denotes the set of all joint distributions $\gamma(x,y)$ whose marginals are $\bmX_{i\delta}$ and $p(\bmX_{i\delta})$, respectively, and $p(\cdot)$ indicates user-specified distributions for the attributes. 
This loss indicates the difference between the distributions of generated attribute values and user-specified distributions $p(\cdot)$ (Bernoulli or normal distribution). 
For instance, if users specify a normal distribution for attributes, we formulate $p(\cdot)$ as follows: 
\begin{align}
\label{eq:q_distribution}
    p(\bmX_{i\delta}) = \mathcal{N}(\frac{\sum_{i=1}^n\bmX_{i\delta}}{n},w^2)
\end{align}
where $\omega$ indicates the user-specified deviation of a normal distribution. 

\subsubsection{Attribute generation}
\label{sssec:attribute generation}
In order to reduce $L^\text{attr}_\text{class feature}$ and $L^\text{attr}_\text{graph feature}$, first 1) we obtain base attribute vectors for nodes by computing the product of $\bmU,\bmV\trans$ so that two nodes in the same class (reflecting the effect of the topology and attributes) should share similar attribute values, 
and then 2) we apply user-specified distribution to the base attribute vectors so that the attribute values should follow the distribution. 

Therefore, GenCAT can generate an adjacency matrix $\bmS$ and an attribute matrix $\bmX$ with user-controlled class structure. Also, GenCAT outputs class labels $\bmC$ of nodes, which are obtained based on class size distribution. 

\subsection{Algorithm}
\label{ssec: algorithm}
In this subsection, we explain the attributed graph generation algorithms of GenCAT in detail. 
Algorithm~\ref{al: our method} describes the whole procedure of graph generation.
It consists of three phases, latent factor generation phase, adjusting proportion phase, and graph generation phase.
First, in latent factor generation phase, we generate and initialize latent factors by graph/class features specified by users. 
Second, in adjusting proportion phase, we adjust the latent factors $\bmU$ and $\bmU'$ to minimize the loss $L_\text{mean}$ (Eq.~\eqref{eq: loss M}) and $\bmV$ to minimize the loss $L^\text{attr}_\text{class feature}$ (Eq.~\eqref{eq:loss_att_class_feature}) for graph generation. 
Finally, in graph generation phase, we generate an adjacency matrix $\bmS$ and an attribute matrix $\bmX$ from the latent factors. 

\begin{algorithm}[!t]
    \caption{Graph generation}
    \label{al: our method}
        \DontPrintSemicolon
            \SetKwInOut{Input}{input}
            \SetKwInOut{Output}{output}
            \SetKwFunction{Expand}{Expand}
            \Input{$n,m,d,k,\bmM,\bmD,\bmH,\phi_{C},\omega,r,f_S,f_X$} 
            \Output{adjacency matrix $\bmS$, attribute matrix $\bmX$, class label ${\bm C}$}
            \textbf{\#\#\# Latent factors generation phase \#\#\#}\\
            $class\_size = power\ law(k,\phi_C)$\\
            \For{$i=1$ to $n$}{
                $\bmC_i$ = $Rand_{int}$(range=($1,k$),weight=$class\_size$)\\
                $\bmU_{i.}$ = $normal (\bmM_{\bmC_i.},\bmD_{\bmC_i.}$)\\
            }
            $\bmU' = \bmU$ \\
            \For{$l=1$ to $k$}{
                \If{$\bmM_{ll}<\frac{1}{k}$}{
                    \For{$i\in\Omega_l$}{
                        $\bmU_{i.} = f_{\rm rev}(\bmU_{i.})$ (Eq.~\eqref{eq: U prime}) \hfill $\vartriangleright$ negative topology type
                    }
                }
            }
            $\bmV$ = $\bmH$\\
            \textbf{\#\#\# Adjusting proportion phase \#\#\#}\\
            \For{$l=1$ to $k$}{
                \If{$\bmM_{ll}\ge\frac{1}{k}$}{
                    $T_{min} = \argmin_T(\|\bmM_{l.} - \frac{1}{|\Omega_l|} \sum_{i\in\Omega_l}f_{\rm AP}(\bmU_{i.},T)*f_{\rm AP}(\bmU_{i.},T)\|_F)$ \\
                    {\nonl\hfill $\vartriangleright$ minimize $L_\text{mean}$ (Eq.~\eqref{eq: loss M}) by using $f_{\rm AP}$ (Eq.~\eqref{eq: adjust})}\\
                    \For{$i\in\Omega_l$}{
                        $\bmU_{i.} = f_{\rm AP}(\bmU_{i.},T_{min})$ \\
                        $\bmU'_{i.}=\bmU_{i.}$
                    }
                }\Else{
                    $T_{min} = \argmin_T(\|\bmM_{l.} - \frac{1}{|\Omega_l|} \sum_{i\in\Omega_l}f_{\rm AP}(\bmU_{i.},T)*f_{\rm rev}(f_{\rm AP}(\bmU_{i.},T))\|_F)$ \\
                    {\nonl\hfill $\vartriangleright$ minimize $L_\text{mean}$ by using $f_{\rm rev}$ (Eq.~\ref{eq: inverseU}) and $f_{\rm AP}$}\\
                    \For{$i\in\Omega_l$}{
                        $\bmU_{i.} = f_{\rm AP}(\bmU_{i.},T_{min})$ \\
                        $\bmU'_{i.}=f_{\rm rev}(\bmU_{i.})$
                    }
                }
            }
            \For{$\delta=1$ to $d$}{
                $T_{min} = \argmin_T(\|\bmH_{\delta.} - \frac{1}{|\Omega_l|} \sum_{i\in\Omega_l}\bmU_{i.}*f_{\rm AP}(\bmV_{\delta.},T)\|_F)$ \\
                {\nonl\hfill $\vartriangleright$ minimize $L^\text{attr}_\text{class feature}$ (Eq.~\eqref{eq:loss_att_class_feature}) by using $f_{\rm AP}$} \\
                $\bmV_{\delta.} = f_{\rm AP}(\bmV_{\delta.},T_{min})$ 
            }
           \textbf{\#\#\# Graph generation phase \#\#\#}\\
            $\bmS$ = $f_S(\bmU,\bmU',n,m,k,r$) \hfill $\vartriangleright$ $f_S =$  Edge\_generation\\
            $\bmX = f_X(\bmU,\bmV,\omega$) \\
            {\bf return} $\bmS,\bmX,\bmC$
\end{algorithm} 
\subsubsection{Latent factor generation}
In the latent factors generation phase (lines $1$--$11$),
we first generate a class size distribution from a power law distribution controlled by input parameter $\phi_C$ (line $2$)\footnote{GenCAT allows users to adopt a normal distribution and to directly input class size distribution ${\bm \rho}$.}. 
Next, we generate a class label $\bmC$ for each node based on the class size distribution and initialize $\bmU$ by a normal distribution whose average is $\bmM$ and deviation is $\bmD$ (lines $3$--$5$). 
To support both positive and negative topology types, we initialize node-class connection proportions, $\bmU'$, by $\bmU$ and then reverse the node-class membership proportions of nodes which are in classes typed with negative topology (lines 6--10). 
Note that this reverse realizes that each node has the large value in the \mem\ proportions for a class that the node belongs to according to Eq.~\eqref{eq: U prime}. 
Then, we initialize attribute-class proportions, $\bmV$, by $\bmH$ so that the attribute-class proportions reflect user-specified correlation between the attributes and classes (line $11$). 

\subsubsection{Adjusting proportion phase}
The goal of this phase is to rescale the \mem\ proportions and attribute-class proportions to reduce the losses imposed by the inputs of class preference mean and \attcor, respectively. 
The reason that we need to adjust these proportions is that the initialization of the proportions is not designed to minimize the losses (Eq. \eqref{eq: loss M} and \eqref{eq:loss_att_class_feature}) since the initialization aims to capture the tendency of user-specified statistics. 

First, we adjust the \mem\ proportions $\bmU$ for each class to minimize the loss $L_\text{mean}$ shown in Eq.~\eqref{eq: loss M} (lines $13$--$23$).
The procedure adjusting $\bmU$ depends on whether a class is typed with positive topology (lines 14--18) or negative topology (lines 19--23). 
As for the positive topology type, we adopt a grid parameter search for $T$ from $0$ to $1$ in $0.05$ step, in order to minimize $L_\text{mean}$ shown in Eq.~\eqref{eq: loss M} (line 15). 
To rescale the \mem\ proportions with probability values for the minimization, we utilize the same idea in \cite{xie2016unsupervised} by adopting rescale function $f_{AP}$ below: 
\begin{eqnarray}
    \label{eq: adjust}
f_{\rm AP}(\bmU_{i.},T) = \bmU_{i.}^{\frac{1}{T}} / \sum_j^k(\bmU_{ij}^{\frac{1}{T}})
\end{eqnarray}
where $i$ is a node ID and $T$ is a parameter which controls the degree of rescale. 
Then, for each node $i$ in the class $l$, we update $\bmU_{i.}$ by using $f_{\rm AP}$ with $T_{min}$ that is the output of the grid search (line 17) and update $\bmU'_{i.}$ to the same value as $\bmU_{i.}$ (line 18). 
As for the negative topology type, we use $f_{\rm rev}$ (Eq.~\eqref{eq: inverseU}) to compute $\bmU'$ from $\bmU$ and minimize $L_\text{mean}$ by utilizing a grid search similarly as positive topology type (line 20). 
Then, for each node $i$ in the class $l$, we update $\bmU_{i.}$ by using $f_{\rm AP}$ with $T_{min}$ (line 22) and update $\bmU'_{i.}$ by using $f_{\rm rev}$ since the class $l$ is typed with negative topology (line 23). 
Note that we adjust $\bmU_{i.}*\bmU'_{i.}$ with $\bmM$ (See Eq.~\eqref{eq: loss M}) as an approximation since the cost to obtain the exact proportion of edges between classes is $O(kn^2)$.\footnote{$O(kn^2)$ stems from the matrix multiplication $\bmU{\bmU'}\trans$ to consider all pairs of nodes.} 

Second, we adjust the attribute-class proportions $\bmV$ for each attribute to minimize the loss $L^\text{attr}_\text{class feature}$ shown in Eq. \eqref{eq:loss_att_class_feature} (lines 25--26). 
We can obtain adjusted $\bmV$ by using a grid search similarly as $\bmU$.

\subsubsection{Graph generation phase}
In the graph generation phase, 
we generate the adjacency matrix $\bmS$ (line $28$) and attribute matrix $\bmX$ (line $29$) by using the adjusted latent factors.

\begin{algorithm}[!t]
     \caption{Edge\_generation($\bmU,\bmU',n,m,k,r$)}
     \label{al: edge construction}
         \DontPrintSemicolon
            \SetKwInOut{Input}{input}
            \SetKwInOut{Output}{output}
            \SetKwFunction{Expand}{Expand}
            \Input{$\bmU,\bmU',n,m,k,r$}
            \Output{$\bmS$}
            $\phi^{min}_d = \argmin_{\phi_d} |m-\sum_{i=1}^n({\rm powerlaw}(n,\phi_d))_i/2 |$ \\
            $\bmtheta = {\rm sort}({\rm powerlaw}(n,\phi^{min}_d)$, \text{descending order}) \\ 
            ${\bm \bmtheta^{\prime}} = [0]^n$\\
            \For{$i=1$ to $n$}{
                $counter = 0$\\
                \While{$counter < r$ and $\bmtheta^{\prime}_i < \bmtheta_i$}{ 
                    \textbf{\#\#\# Class selection from source node \#\#\#}\\
                    $Z = \{\}$\\
                    \For{$iter=1$ to $\bmtheta_i - \bmtheta_i'$}{
                        $Z = Z \bigcup Rand_{int}($range$=(1,k), $weight$=\bmU_{i.})$
                    }
            \textbf{\#\#\# Target node selection from class \#\#\#}\\
            \For{$l \in Z$}{
                $j = Rand_{int}($range$=(1,n),$weight$={\bmU'}\trans_{.l})$\\
                \If{$\bmS_{ij} == 0$ and $\bmtheta^{\prime}_i < \bmtheta_i$ and $\bmtheta^{\prime}_j < \bmtheta_j$}{
                    $\bmS_{ij} = 1$\\ 
                    $\bmS_{ji} = \bmS_{ij}$ \hfill $\vartriangleright$ undirected graph\\
                    $(\bmtheta^{\prime}_i,\bmtheta^{\prime}_j) = (\bmtheta^{\prime}_i+1,\bmtheta^{\prime}_j+1)$ \hfill $\vartriangleright$ increment degrees
                    }
                }
            $counter = counter + 1$
            }
        }
        {\bf return} $\bmS$
\end{algorithm}

\smallskip
\noindent{{\bf Edge generation.}}
Algorithm~\ref{al: edge construction} describes how adjacency matrix $\bmS$ is generated.
First, we adopt a grid parameter search for a power law parameter $\phi^{min}_d$ from $1$ to $3$ in $0.01$ step, so as to reduce the loss between the number of edges and the sum of the expected node degree proportions\footnote{Users can input an arbitrary node degree distribution. 
In our algorithm, we show a case of using a power law distribution that node degrees in real-world graphs often follow.} (line $1$). 
Then, ${\bm \theta}$ is generated by using a power law distribution with $\phi^{min}_d$, and we sort it in descending order so that we start edge generation from high degree nodes as we mentioned in Section~\ref{sssec: graph generation} (line $2$). 
Let ${\bm \theta'}$ be the actual node degrees during the edge generation.
It is initialized as zeros (line $3$) and its entries ${\bm \theta'_i}$ and ${\bm \theta'_j}$ are incremented when a new edge ($i$, $j$) is generated (line $17$).
For each node $i$, edges are iteratively generated until $\bmtheta'_i$ gets close to the expected node degree proportion ${\bmtheta_{i}}$ by {\it Class selection from source node} step (lines $8$--$10$) and {\it Target node selection from class} step (lines $12$--$17$). 
\seiji{Thanks to these steps, GenCAT generates edges according to the stochastic process specified by $\bmU$ and ${\bmU'}\trans$. }
In {\it Class selection from source node} step, classes are chosen from the \mem\ proportions of source node $i$, $\bmU_{i.}$ (line $10$). 
{\it Target node selection from class} step is iterated for the classes selected in the previous step (line $12$). 
We select target node $j$ from the node-class connection proportions ${\bmU'}\trans$ (line $13$). 

In order to accelerate {\it Target node selection from class} step, we generate a list $to\_node\_prob$ by which we can obtain a node ID based on the probability of selecting the node, in a similar way as the inverse CDF of ${\bmU'}\trans$. 
Setting a step size $w$, which balances memory size and the accuracy of the selection, $to\_node\_prob$ is formulated below:
\begin{equation}
  to\_node\_prob[c] = f_{CDF}^{-1}(w\cdot c)
\end{equation}
where $c \in \mathbb{N} $ and $w*c \le 1$. 
The length of $to\_node\_prob$ is $1/w$. 
We set $c$ by the following equation\footnote{If node IDs are numbered from $0$ to $n-1$, we just replace the ceiling function to the floor function.}:
\begin{equation}
  c = \lceil Rand(range(0,1)) / w \rceil
\end{equation}
We set $w=1/(100n)$ so as to vary with $n$ since there is a trade-off between memory size\footnote{$to\_node\_prob$ needs the space complexity, $O(kn)$, which is one of the largest elements (The details are described in Section~\ref{sssec: space complexity}).} and the accuracy of the selection. 

If neither the actual node degrees $\bmtheta'_i$ or $\bmtheta'_j$ reach the expected node degrees $\bmtheta_i$ and $\bmtheta_j$, respectively\footnote{If the actual node degrees reach the expected node degrees, that means the node has enough number of edges.}, we generate an edge between them and then increment their node degrees (lines $14$--$17$). 
Note there is a possibility that the edge generation loop (lines 6--18) does not stop because there is such a case that the last node of the loop cannot have its expected degree. 
To avoid this, we exit the loop at the user-specified $r$ iterations (line $6$)\footnote{Although we adjust the sum of all the node degrees in ${\bm \theta}$ to be the number of edges $m$, some candidate edges may not be generated when the node degrees of the adjacent nodes exceeds the expected ones, so the actual number of the generated edges tends to be smaller than the expected number of edges, $m$.}. 
It is our future work to guarantee the theoretical quality bounds of the generated graphs.

\smallskip
\noindent{{\bf Attribute generation.}}
Attribute matrix $\bmX$ is generated by $f_X(\bmU,\bmV,\omega)$ (line $29$ in Algorithm~\ref{al: our method}). 
As we described in Section \ref{sssec:attribute generation}, we obtain base attribute vectors by multiplying $\bmU\bmV\trans$ so that nodes in the same class should share similar attribute values, 
and then we apply user-specified distribution to the base attribute vectors so that the attribute values should follow the distribution in order to reduce the loss $L^\text{attr}_\text{graph feature}$ shown in Eq.~\eqref{eq:loss_att_feature}. 
As for the application of user-specified distributions, GenCAT supports two types of distributions, normal distribution and Bernoulli distribution. 
First, if users specify a normal distribution for attributes, we calculate $\bmX_{i\delta}=(\bmU\bmV\trans)_{i\delta} + \mathcal{N}(0,\omega^2)$.
We normalize all attribute values to [0,1] without the loss of generality. 
Second, if users specify the Bernoulli distribution for attributes, we calculate $\bmX_{i\delta}= \mathcal{B}((\bmU\bmV\trans)_{i\delta})$, where $\mathcal{B}(x)$ is a function which returns $1$ with probability $x$ or $0$ with probability $1-x$. 
It is our future work to support other distributions for the attributes such as power-law and categorical distributions. 

\subsection{Parameter extracting from given graph datasets}
\label{ssec: reproduction}
Recall the second scenario that we described in Section~\ref{sec: problem statement}: we extract parameters from input graphs with class labels and reproduce graphs similar to the input graphs. 
GenCAT uses a parameter extracting function that obtains topology statistic and class features of given graphs; node degrees, class preference mean, class preference deviation, and class sizes. 
Finally, we construct graphs in the same way as the first scenario. 

Thanks to the parameter extracting function, GenCAT easily generates graphs similar to those that users input. 
Additionally, it enables generating graphs in arbitrary size with similar class features of the given graphs by permitting users to change the numbers of nodes and edges.

\subsection{Complexity}
\label{ssec: complexity}
We discuss the time/space complexities of GenCAT.
As is typical in network analytics, 
we focus on sparse graphs~\cite{newman2010intro} 
as real-world graphs are often sparse.
On the sparse condition, the mean of the node degree, $\theta_{Avg}$, can be treated as a constant. 
As the result, $m \propto n$ holds. 
Hence, $m$ is considered to be a much smaller value than $n^2$. 

\subsubsection{Time complexity.}
We analyze the time complexity of the latent factor generation, the edge generation, and the attribute generation, respectively.
First, the complexity for generating the node-class membership/connection proportions and the attribute-class proportions is $O(kn+dk)$ based on their matrix sizes.
Second, adjusting proportion phase consists of two parts, adjusting the node-class membership/connection proportions and adjusting the attribute-class proportions. 
The former phase requires $O(kn)$ to adjust them. 
The latter phase requires $O(dkn)$ to calculate $\bmU\bmV\trans$. 
In practical cases, we can rewrite $\frac{1}{|\Omega_l|}\sum_{i\in\Omega_l}\bmU\bmV\trans$ as $\bmP\bmV\trans$, where $\bmP\inmathbbr{k}{k}$ and $\bmP_{l.} = \frac{1}{|\Omega_l|}\sum_{i\in\Omega_l}\bmU_{i.}$. 
The computational cost of this phase is $O(kn+dk^2)$ owing to this transformation. 
Note that $k$ is typically much smaller than $n$. 

The edge generation consists of the {\it Class selection from source node} step and the {\it  Target node selection from class} step.
In the former step, classes are chosen based on the \mem\ proportions $\bmU$ for each node. 
The cost is $O(kn\theta_{Avg})$ since we select as many classes as the remained node degrees. 
In the latter step, we select a node for each class selected in the former step, based on the transpose of the topology proportions $\bmU'$. 
We generate list {\it to\_node\_prob} to accelerate this selection. 
The cost of this step is $O(1)$ since this step only includes the operation of a random value generation and list access. 
These two steps require $O(mkr)$ since $m=n\theta_{Avg}$ and $r$ is a constant number of iterations for the edge generation. 
Finally, in the attribute generation, the matrix multiplication $\bmU\bmV\trans$ requires $O(dkn)$ and the application of user-specified distribution requires $O(dn)$. 
Hence, the complexity of the attribute generation is $O(dkn)$. 
Therefore, the total time complexity is $O(mkr+dkn)$.

\subsubsection{Space complexity.}
\label{sssec: space complexity}
The largest concern is the adjacency matrix $\bmS$ since the size of $\bmS$ could be as large as $n^2$. 
Hence, we use sparse representation for $\bmS$, such as an adjacency list, with size $O(m)$. 
The sizes of the latent factors $\bmU,\bmU',$ and $\bmV$ are $O(kn), O(kn),$ and $O(dk)$, respectively. 
To utilize inverse transform sampling, we construct lists whose size is $O(kn)$. 
The size of an attribute matrix $\bmX$ is $O(dn)$. 
Therefore, the space complexity is $O(m+kn+dn)$ since $k$ is smaller than $n$.

\subsection{Simulating existing generators}
\label{ssec: simulating}
We show that GenCAT is a general generator of existing methods, LFR and DC-SBM. 
First, LFR specifies the edge fraction of intra- and inter-edges by a mixing parameter. 
So, GenCAT can simulate LFR by adding three conditions that 1) nodes in a class have the same \mem\ proportions: $\bmU_{i.}=\bmU_{j.}$ if $\bmC_i=\bmC_j$, 2) the connection proportions between each class and other classes are the same: $\bmU_{ih_1}=\bmU_{ih_2}$ if $h_1,h_2\ne\bmC_i$, and 3) all classes have the same edge fraction of intra- and inter-edges: $\bmU_{i\bmC_i}=\bmU_{j\bmC_j}$ for all pairs of nodes $(i,j)$. 
To realize the first condition, we can set all elements of $\bmD$ to zero so that every node in a class has the same \mem\ proportion. 
As for the second condition, a given class preference mean should satisfy $\bmM_{lh_1}=\bmM_{lh_2}$ if $h_1,h_2\ne l$ for each class $l$. 
For the third condition, a given class preference mean should also satisfy $\bmM_{ll}=\mu$ where $1\le l\le k$ and $\mu$ is a mixing parameter of LFR. 

Next, DC-SBM specifies the connection proportions between classes. 
So, the edge probabilities of GenCAT and DC-SBM correspond by adding a condition that nodes in a class have the same \mem\ proportions: $\bmU_{i.}=\bmU_{j.}$ if $\bmC_i=\bmC_j$. 
To realize the condition, we can set all elements of $\bmD$ to zero in addition to the first condition of LFR. 
As for graph features, GenCAT can utilize the same distributions as the existing generators. 
Hence, GenCAT can simulate the existing generators in terms of both graph/class features.

\section{Experiments}
\label{sec:experiments}
We next describe an experimental study of GenCAT. Our goal in the experiments is to answer the following questions:
\begin{description}
    \item [Q1] Does GenCAT support users to flexibly control graph/class features regarding \textbf{topology}? (Sec. \ref{ssec:Q1})
    \item [Q2] Does GenCAT support users to flexibly control graph/class features regarding \textbf{attributes}? (Sec. \ref{ssec:Q2})
    \item [Q3] How well does GenCAT scale? (Sec. \ref{ssec:Q3})
    \item [Q4] How precisely does GenCAT reproduce real-world graphs? (Sec. \ref{ssec:Q4})
    \item [Q5] How efficient and effective are the proposed techniques on edge generation? (Sec. \ref{ssec:Q5})
\end{description}

In Q1 and Q2, we validate that GenCAT can output attributed graphs following graph/class features specified by given parameters. 
In Q3 and Q4, we evaluate the performance of scalability and reproducibility of GenCAT compared with existing graph generators.
In Q5, we demonstrate the efficiency and effectiveness of the proposed techniques by conducting an ablation study. 
We use LFR\footnote{\url{https://networkx.github.io/documentation/networkx-2.1/reference/algorithms/generated/networkx.algorithms.community.community_generators.LFR_benchmark_graph.html}} and DC-SBM\footnote{Since the source code of SBM is available, we extend SBM to DC-SBM that incorporates node degrees into its edge generation algorithm. 
Note that we make minor changes to the code of SBM for the extension, but the cost of time and space for SBM is the same as for DC-SBM. \url{https://networkx.org/documentation/stable/reference/generated/networkx.generators.community.stochastic_block_model.html}} as the main competitors because LFR and DC-SBM generate graphs with class labels and controlled class-level connection proportions, which are closest to the capabilities of GenCAT in terms of generated topology structure. 
Since schema-driven (e.g., pgMark and TrillionG) and clustering method-driven (e.g., ANC) approaches generate quite different graphs, it is hard to fairly compare them with GenCAT, so we do not compare them with GenCAT. 
For evaluating scalability, we compare GenCAT with LFR and DC-SBM.
For evaluating reproducibility, we compare GenCAT with LFR, DC-SBM, VGAE\footnote{\url{https://github.com/zfjsail/gae-pytorch}}, and NetGAN\footnote{\url{https://github.com/danielzuegner/netgan}}. 
VGAE is a baseline method of network embedding approaches. 
NetGAN is the-state-of-the-art deep learning-based graph generator. 
We do not compare with GraphVAE as it can only be used on very small graphs. 

GenCAT is implemented in Python$3$. 
The experiments are operated on Intel(R) Xeon(R) CPU E5-2640 v4 @ 2.40GHz with 1TB memory. 
All experiments are operated on a single thread and a single core. 

\subsection{Q1: Does GenCAT support users to flexibly control graph/class features regarding topology?}
\label{ssec:Q1}
We here validate that GenCAT generates graphs following the users' given graph/class features regarding the topology. 

\begin{table}[t]
  \caption{MAPE between the expected node degree $\theta$ and the actual node degree $\theta'$.}
  \label{tb: degree}
  \begin{center}
  \setlength{\tabcolsep}{3pt}
	\begin{tabular}{l|ccccc} 
        $m$ & $2^{16}$ & $2^{17}$ & $2^{18}$ & $2^{19}$ & $2^{20}$ \\ \hline
        Error & $1.03e$--$3$ & $7.37e$--$4$ & $5.76e$--$4$ &$5.25e$--$4$ & $5.00e$--$4$ \\ 
 	\end{tabular}
 \end{center}
\end{table}
\subsubsection{Graph feature regarding topology}
We show that GenCAT can generate graphs that almost follow given node degrees by calculating the loss of graph features, the difference between the expected node degree ${\bm \theta}$ and the actual node degrees $\bm \theta'$ by Eq.~\eqref{eq: L degree}. 
We vary the number of edges $m$ within the range of $\{2^{16},2^{17},2^{18},2^{19},2^{20}\}$. 
We set the parameters $k,r,n$ as $5,50,m/16$, respectively, all the diagonal elements of $\bmM$ to $0.4$ and the other elements to $0.15$. 
We generate five graphs and report the average of their results. 
Table~\ref{tb: degree} shows the MAPE of node degrees. 
This result shows that the graphs generated by GenCAT almost follows given node degrees. 

\begin{figure}[t]
  \centering
     \begin{minipage}{0.49\hsize}
      \centering
          \includegraphics[trim={0mm 0mm 0mm 0}, width=4cm]{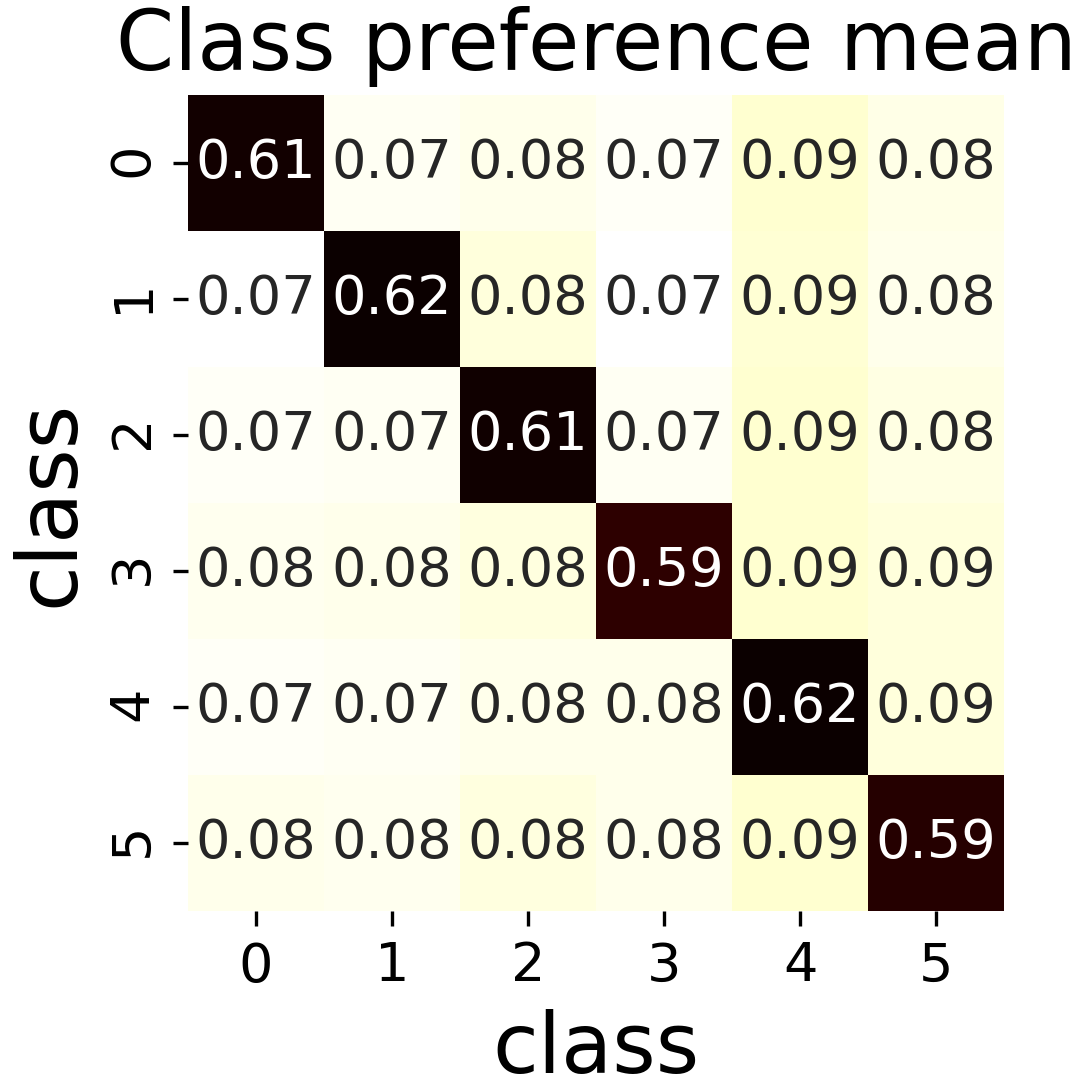}
          \subcaption{Each cell represents the class preference mean of the generated graph, $\bmM^{gen}$. The diagonal elements of $\bmM$ are set to $0.6$ and other elements are set to $0.08$.}
          \label{fg: GenCAT cpm}
     \end{minipage}
     \begin{minipage}{0.49\hsize}
      \centering
          \includegraphics[trim={0mm 0mm 0mm 0}, width=4cm]{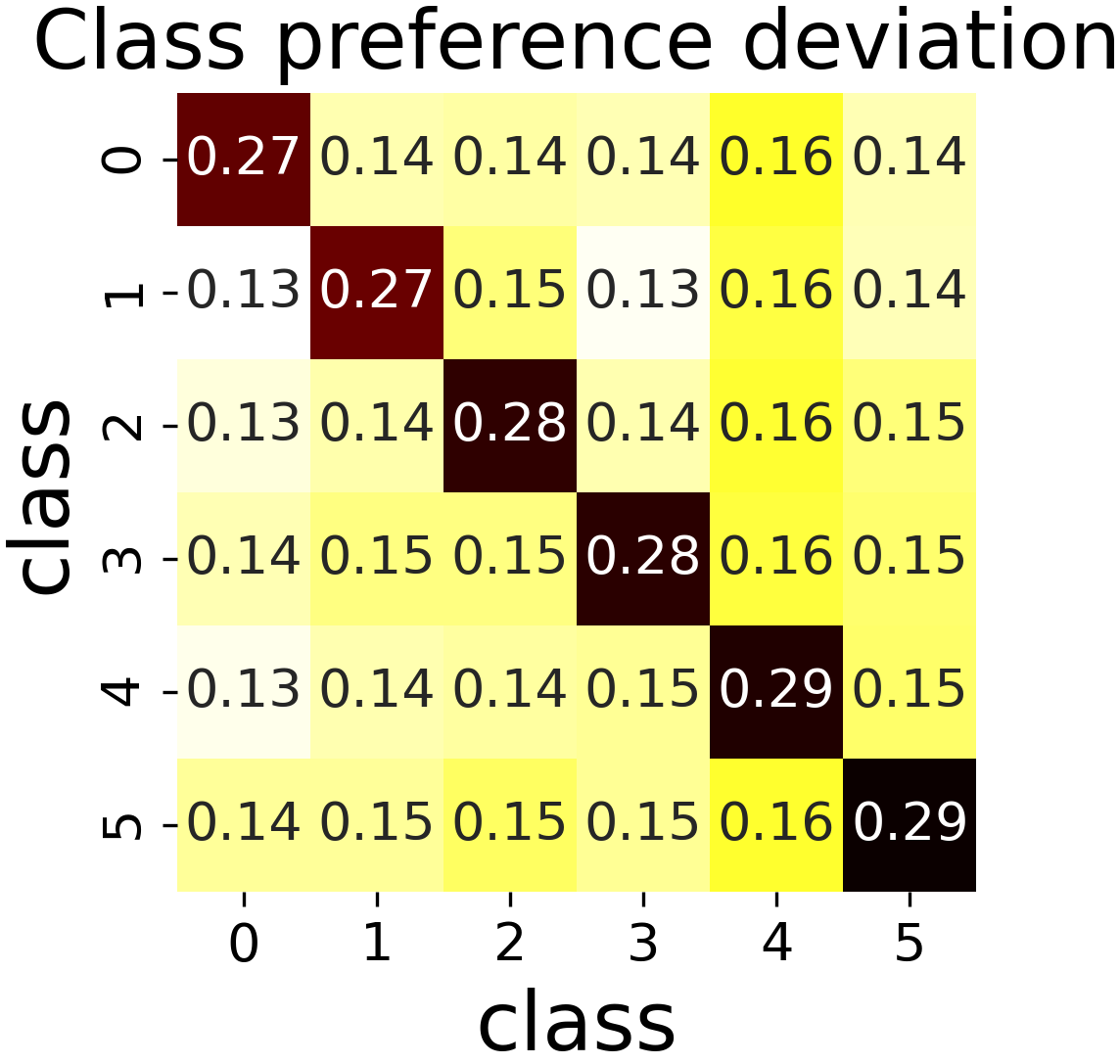}
          \subcaption{Each cell represents the class preference deviation of the generated graph, $\bmD^{gen}$. The diagonal elements of $\bmD$ are set to $[0.2,0.2,0.25,0.25,0.3,0.3]$ and other elements are set to $0.05$.}
          \label{fg: GenCAT cpd}
     \end{minipage}
    \caption{Visualization of class preference mean and class preference deviation. The parameters are set as follows: $n=2^{16}$, $m=2^{20}$, $k=6$.}
  \label{fg: cpm cpd}
\end{figure}
\subsubsection{Class feature regarding topology}
We demonstrate that GenCAT flexibly controls class features in generated graphs. 
Figure~\ref{fg: cpm cpd} shows the heatmaps of the class preference mean and the class preference deviation of the generated graph. 
In this experiment, We set all the diagonal elements of $\bmM$ to $0.6$. 
Also, we set the diagonal elements of $\bmD$ to $[0.2,0.2,0.25,0.25,0.3,0.3]$, respectively, and the other elements to $0.05$. 
Let $\bmM^{gen}$ and $\bmD^{gen}$ be a class preference mean and a class preference deviation of the generated graph. 
First, we demonstrate that GenCAT can control the class-level connection proportions. 
Figure~\ref{fg: GenCAT cpm} shows that the diagonal elements of $\bmM^{gen}$ are almost $0.6$. 
The reason that the diagonal elements of $\bmM^{gen}$ do not completely match $0.6$ is that the edge generation step is based on the probabilistic procedures. 
This figure validates that GenCAT supports classes typed with positive topology. 
Figure~\ref{fg: GenCAT cpd} shows that GenCAT can control the deviation of the connection proportions between nodes and classes. 
There is a tendency that $\bmD^{gen}_{00}$ and $\bmD^{gen}_{11}$ are smaller than $\bmD^{gen}_{44}$ and $\bmD^{gen}_{55}$. 
The reason that the diagonal elements of $\bmD^{gen}$ do not completely match the values we set is that GenCAT prioritizes other constraints higher than estimating class preference deviation in the edge generation, such as the node degrees and the class preference mean. 
This result indicates that GenCAT supports the deviation of the connection proportions, albeit at a lower priority.
In summary, the experiments validate that GenCAT supports the connection proportions between nodes and classes by using the class preference mean and the class preference deviation. 
We compare GenCAT with other generators in Section~\ref{ssec:Q4}. 

\begin{figure}[t]
  \centering
     \begin{minipage}{0.49\hsize}
      \centering
          \includegraphics[trim={0mm 0mm 0mm 0}, width=4. cm]{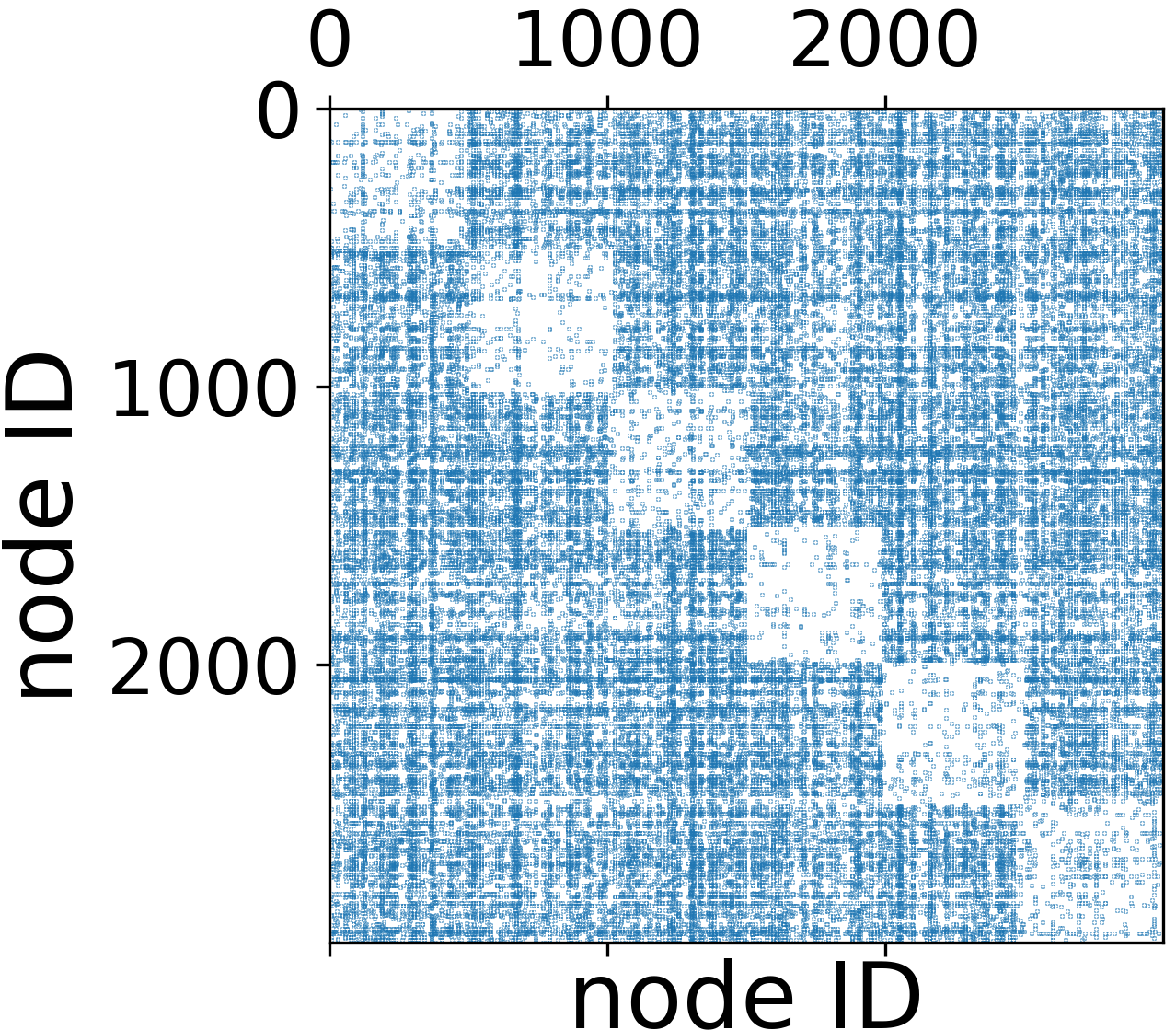}
          \subcaption{Negative topology classes. All diagonal elements of $\bmM$ are set to $0.05$.}
          \label{fg: heterophily}
     \end{minipage}
     \begin{minipage}{0.49\hsize}
      \centering
          \includegraphics[trim={0mm 0mm 0mm 0}, width=4. cm]{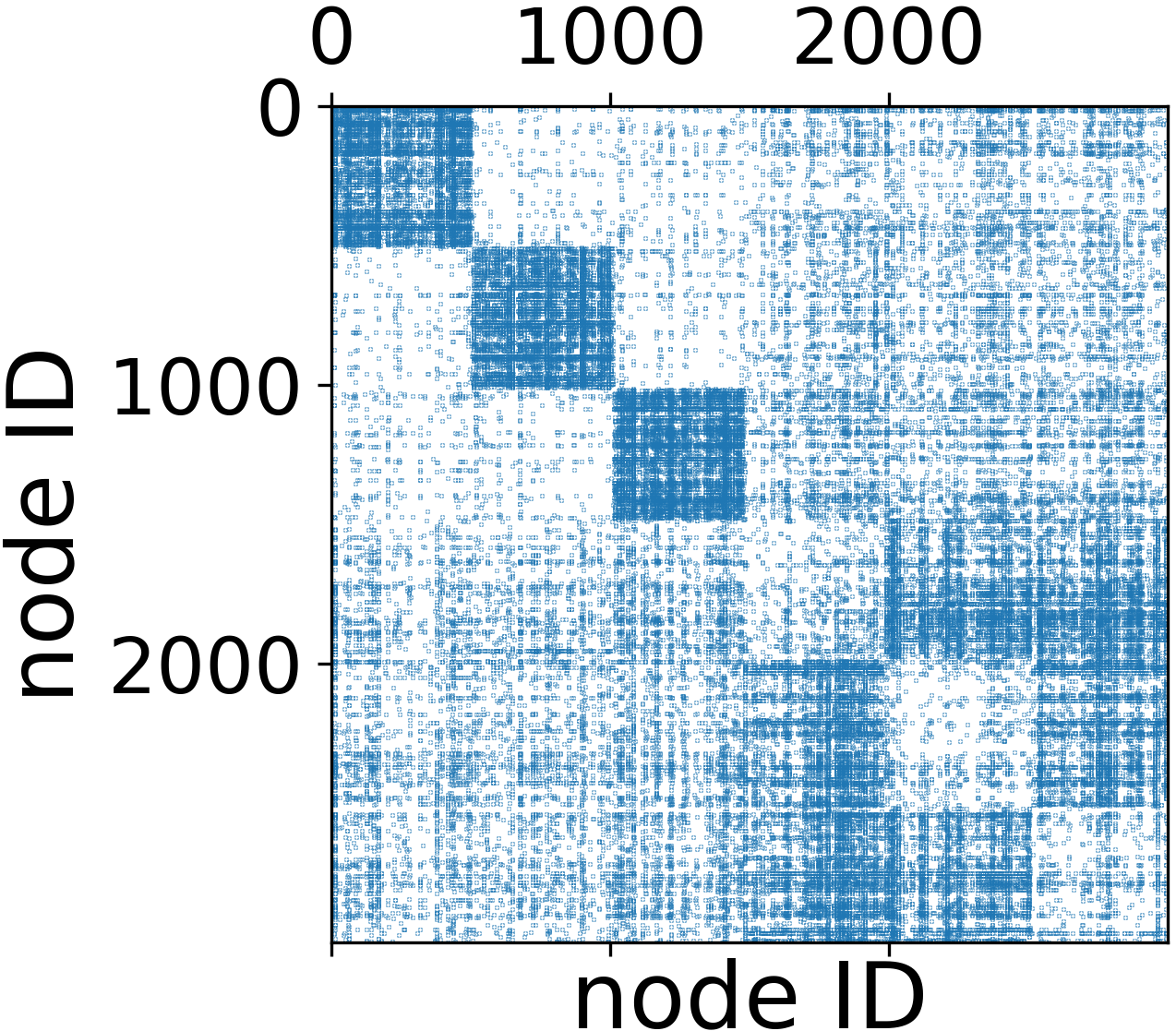}
          \subcaption{Coexistence of positive and negative topology classes. The diagonal elements of $\bmM$ of class $0,1,2$ are set to $0.9$ and the ones of class $3,4,5$ are set to $0.05$.}
          \label{fg: homo hetero}
     \end{minipage}
    \caption{Visualization of adjacency matrices. The parameters are set as follows: $n=3000$, $m=30000$, $k=6$.}
  \label{fg: compatibility}
\end{figure}
Next, we also validate that GenCAT supports various types of class labels with negative and mixed topology types in graphs, respectively. 
Figure~\ref{fg: compatibility} shows the adjacency matrices of the generated graphs. Blue dots indicate that there exist edges between nodes whose identifiers are the node IDs on vertical and horizontal axes. 
In this experiment, we set the number of classes to six, and we can observe the classes as blocks in the diagonal part of the figure. 
Figure~\ref{fg: heterophily} additionally shows the case that all classes are negative types. 
We set all diagonal elements of $\bmM$ to $0.05$, lower than the average. 
The nodes clearly tend to connect with nodes from other classes. 
We demonstrate a more complicated case that both positive and negative topology types in a single graph, in Figure~\ref{fg: homo hetero}. 
Nodes from class $0,1,2$ are densely connected inside since the types of their classes are positive topology. 
Nodes from class $3,4,5$ tend to connect with nodes from other classes. 
In summary, we confirm that GenCAT flexibly controls the connection proportions between nodes and classes.

\subsection{Q2: Does GenCAT support users to flexibly control graph/class features regarding attributes?}
\label{ssec:Q2}
We demonstrate that GenCAT can generate node attributes following the users' given graph/class features regarding the attributes. 
In this experiment, we set the parameters $d,k,n$ as $2,4,5000$, respectively\footnote{To make visualization easier to understand, we set the number of attributes $d$ to $2$. Note that GenCAT can efficiently generate more attributes since its time and space complexities are linear to $d$ (see Section~\ref{ssec: complexity})}, 
all the diagonal elements of $\bmM$ to $0.7$, the other elements to $0.1$, the diagonal elements of $\bmD$ to $[0.2,0.2,0.25,0.25]$, respectively, and the other elements of $\bmD$ to $0.1$. 
We set a normal distribution as the distribution of the attribute and its deviation $\omega$ to $0.2$. 
We vary attribute-class correlations $\bmH$ in three patterns of separation degrees between classes: $\bmH^{(1)}=[[0.5,0.0,0.0,0.5],$ $[0.0,0.5,0.0,0.5]]$, $\bmH^{(2)}=[[0.4, 0.1, 0.1, 0.4],$ $[0.1, 0.4, 0.1, 0.4]]$, and $\bmH^{(3)}=[[0.3, 0.2, 0.2, 0.3],$ \\ $[0.2, 0.3, 0.2, 0.3]]$. 

\begin{figure}[t]
  \begin{minipage}{0.32\hsize}
      \centering
  \includegraphics[trim={0mm 0mm 0mm 0},width=3.8cm]{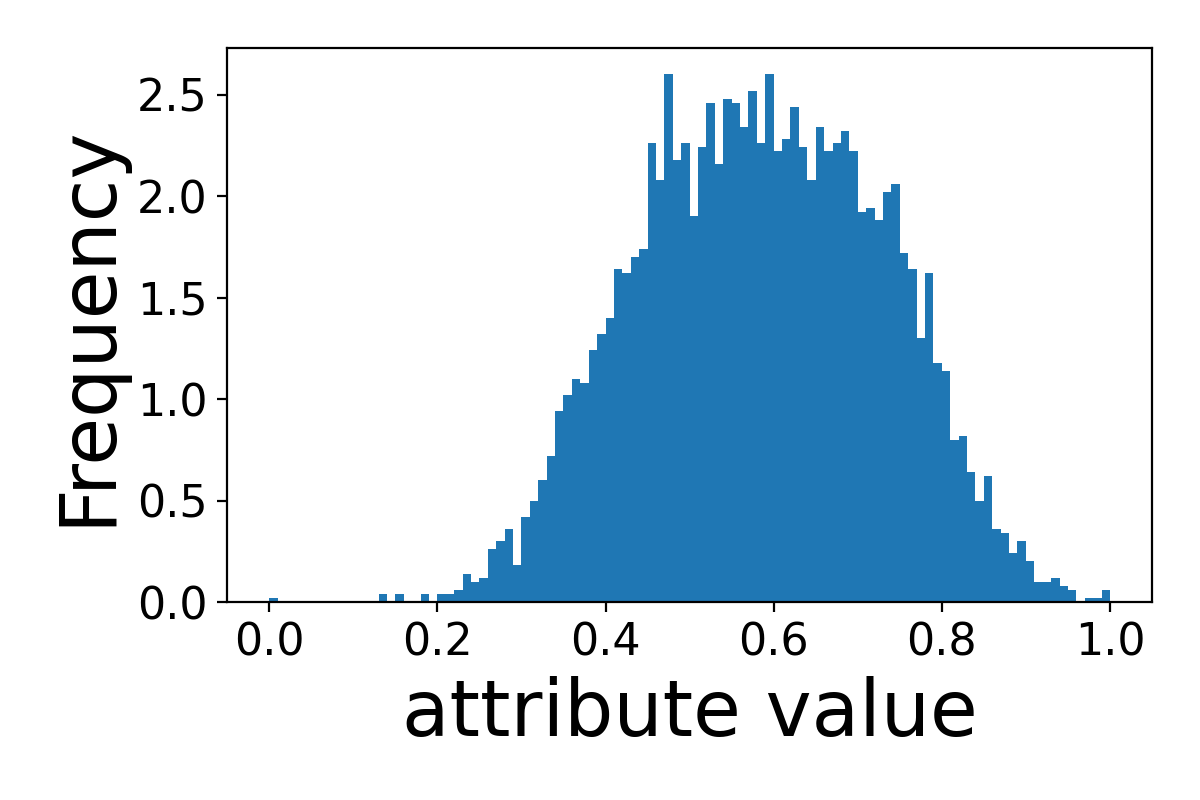}
  \subcaption{$\bmH_{1.}^{(1)}=[0.5,0.0,0.0,0.5]$.}
  \label{fg:04_att_hist_05}
 \end{minipage}
  \begin{minipage}{0.32\hsize}
      \centering
  \includegraphics[trim={0mm 0mm 0mm 0},width=3.8cm]{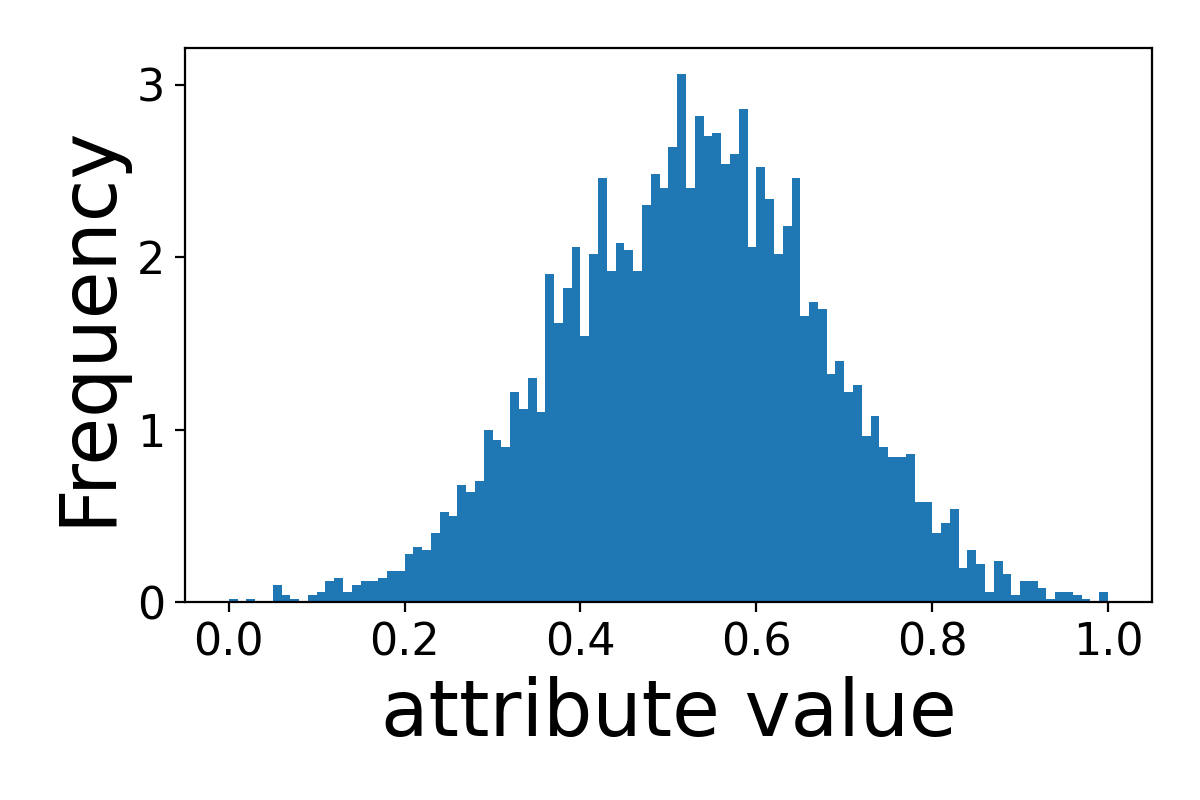}
  \subcaption{$\bmH_{1.}^{(2)}=[0.4, 0.1, 0.1, 0.4]$.}
  \label{fg:04_att_hist_04}
 \end{minipage}
   \begin{minipage}{0.32\hsize}
      \centering
  \includegraphics[trim={0mm 0mm 0mm 0},width=3.8cm]{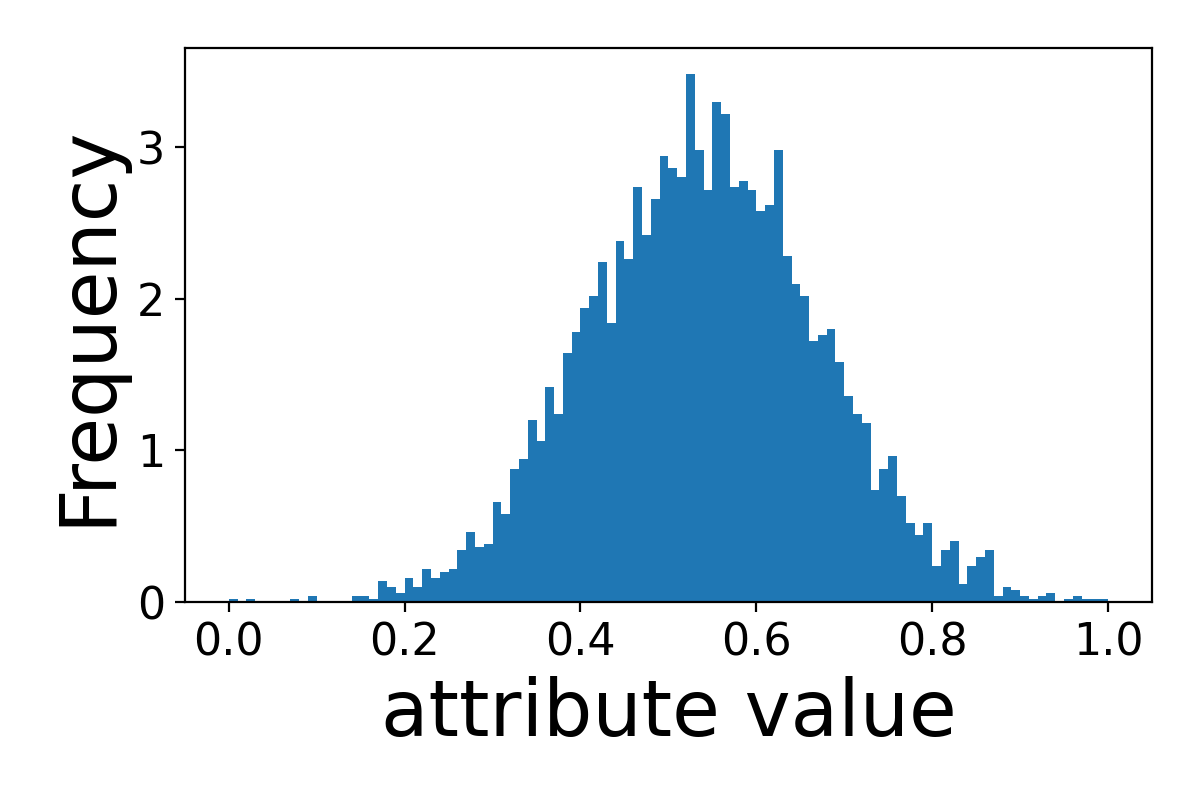}
  \subcaption{$\bmH_{1.}^{(3)}=[0.3, 0.2, 0.2, 0.3]$.}
  \label{fg:04_att_hist_03}
 \end{minipage}
  \caption{Histograms of the generated attributes (attribute$1$ in Figure~\ref{fg:04_Att_plot}) with three variations of attribute-class correlations, $\bmH^{(1)},\bmH^{(2)},\bmH^{(3)}$. 
  All attributes of the three variations follow user-specified distributions which are normal distributions even when given attribute-class correlations are different from each other. 
  }
  \label{fg:04_Att_hist}
\end{figure}
\begin{table}[t]
  \caption{Comparison of the distributions of the generated attributes and the matrix multiplication $\bmU\bmV\trans$ by using the EM distance. 
  $D_{\rm EM}(\bmX,p(\bmX))$ denotes the sum of the EM distance between $\bmX$ and $p(\bmX)$ for all attributes, as shown in Eq. \eqref{eq:loss_att_feature}. 
  $\bmH^{(1)}$, $\bmH^{(2)}$, and $\bmH^{(3)}$ are the variations of given attribute-class correlations. }
  \label{tb:04_EMD_att}
  \begin{center}
  \setlength{\tabcolsep}{3pt}
	\begin{tabular}{l|ccc} 
        EM distance & $\bmH^{(1)}$ & $\bmH^{(2)}$ & $\bmH^{(3)}$ \\ \hline
        $D_{\rm EM}(\bmX,p(\bmX))$ & $0.049$ & $0.089$ & $0.108$ \\ 
        $D_{\rm EM}(\bmU\bmV\trans,p(\bmX))$ & $0.473$ & $0.492$ & $0.481$ 
 	\end{tabular}
 \end{center}
\end{table}
\subsubsection{Graph feature regarding attribute}
First, in order to validate that the attributes of generated graphs with various attribute-class correlations follow user-specified distributions (i.e., the graph feature regarding attributes), Figure~\ref{fg:04_Att_hist} shows the histograms of a single attribute $\bmH_{1.}$ for each variation of \attcor s. 
We observe that all distributions of the attribute values in the generated graphs with $\bmH^{(1)}$, $\bmH^{(2)}$, and $\bmH^{(3)}$ follow normal distributions, which are shown in Figure~\ref{fg:04_att_hist_05}, \ref{fg:04_att_hist_04}, and \ref{fg:04_att_hist_03}, respectively. 
Hence, we conclude that GenCAT can generate attribute values according to the given distribution even when users input various \attcor s. 
To quantitatively evaluate the effectiveness of the application of distributions, we investigate the distance between the distributions of generated attributes $\bmX$ and user-specified distributions $p(\bmX)$ by using the EM distance (Eq.~\eqref{eq:loss_att_feature}). 
We also calculate the distance between user-specified distributions and the distributions of the matrix multiplication $\bmU\bmV\trans$ (i.e., we directly use the matrix multiplication) for comparison. 
Table \ref{tb:04_EMD_att} shows the results of the EM distances for $\bmH^{(1)}$, $\bmH^{(2)}$, and $\bmH^{(3)}$. 
Since the distances between $\bmX$ and $p(\bmX)$ are smaller than the distance between $\bmU\bmV\trans$ and $p(\bmX)$, we confirm the effectiveness of the application of user-specified distributions in the attribute generation. 

\begin{figure}[t]
  \begin{minipage}{0.32\hsize}
      \centering
  \includegraphics[trim={0mm 0mm 0mm 0},width=3.8cm]{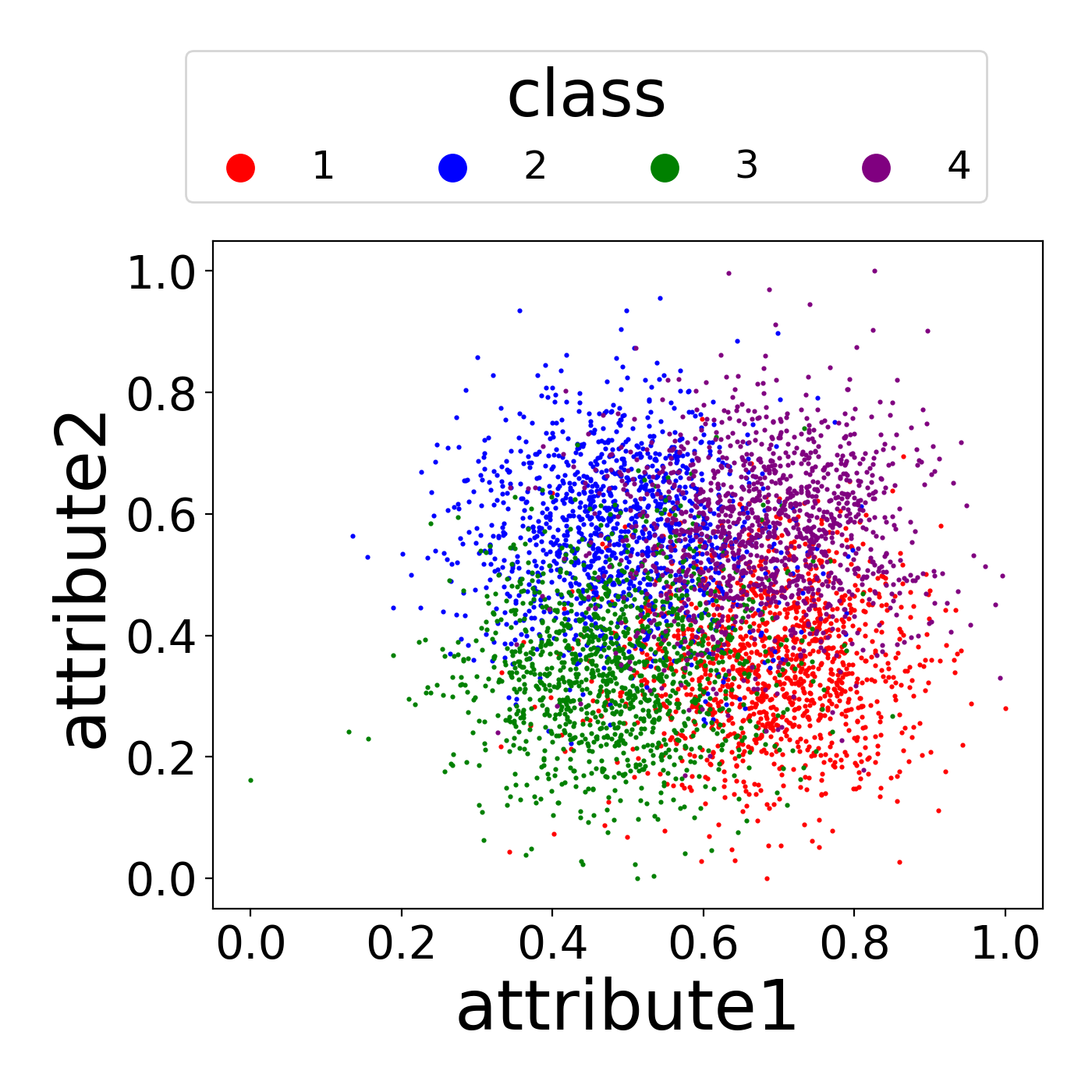}
  \subcaption{$\bmH^{(1)}=[[0.5,0.0,0.0,0.5],$ $[0.0,0.5,0.0,0.5]]$.}
  \label{fg:04_att_plot_05}
 \end{minipage}
  \begin{minipage}{0.32\hsize}
      \centering
  \includegraphics[trim={0mm 0mm 0mm 0},width=3.8cm]{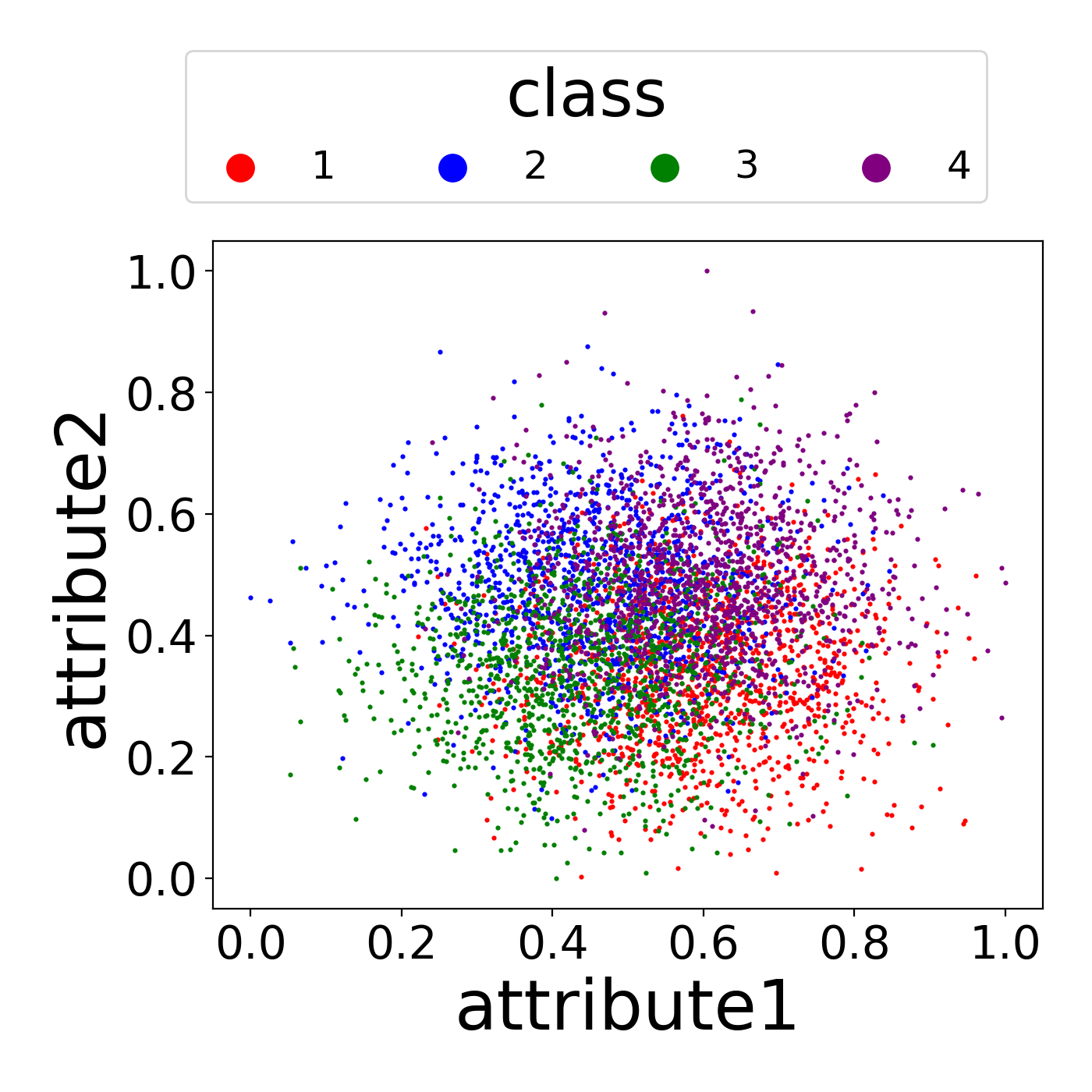}
  \subcaption{$\bmH^{(2)}=[[0.4,0.1,0.1,0.4],$ $[0.1,0.4,0.1,0.4]]$.}
  \label{fg:04_att_plot_04}
 \end{minipage}
   \begin{minipage}{0.32\hsize}
      \centering
  \includegraphics[trim={0mm 0mm 0mm 0},width=3.8cm]{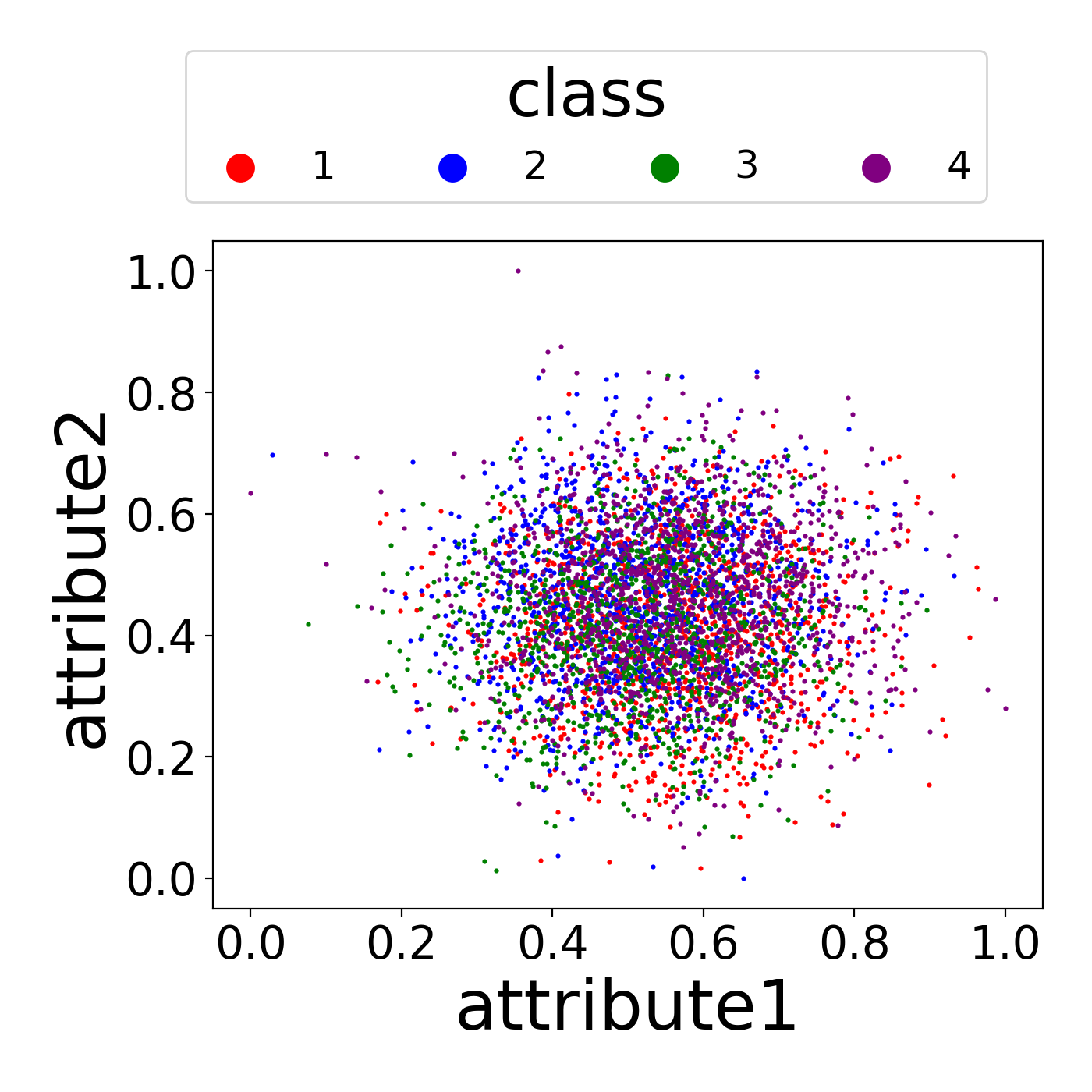}
  \subcaption{$\bmH^{(3)}=[[0.3,0.2,0.2,0.3],$ $[0.2,0.3,0.2,0.3]]$.}
  \label{fg:04_att_plot_03}
 \end{minipage}
  \caption{Distributions of the generated attributes with three variations of attribute-class correlations. We depict the $2$-D plots of the values of two attributes. The colors indicate classes which nodes belong to.}
  \label{fg:04_Att_plot}
\end{figure}
\subsubsection{Class feature regarding attribute}
Next, to show that the generated attributes support the class features specified by users, Figure~\ref{fg:04_Att_plot} depicts 2-D plots of the values of two attributes in the generated graphs with $\bmH^{(1)}$, $\bmH^{(2)}$, and $\bmH^{(3)}$. 
Intuitively, a range of values of $\bmH$ indicates the differences of the attribute values for classes.
Since $\bmH^{(1)}$ has a wide range of values (i.e., $[0.0,0.5]$) the attributes in different classes tend to have dissimilar values in Figure~\ref{fg:04_att_plot_05}. 
Then, Figure~\ref{fg:04_att_plot_04}, \ref{fg:04_att_plot_03} show that the attributes of the nodes in classes are more mixed when the values of the attribute-class correlations are more similar between classes. 
Through this experiment, we observe that GenCAT can flexibly control the class structure in the generated attributes by user-specified attribute-class correlations. 
In summary of this subsection, we show that the attributes in graphs generated by GenCAT closely follow the users' desired graph/class features. 

\subsection{Q3: How well does GenCAT scale?}
\label{ssec:Q3}
\begin{table*}[t]
  \caption{Execution time and memory consumption: TO and OOM indicate that the execution is not finished in $36$ hours and is out of memory, respectively.}
  \label{tb: complexity}
  \begin{center}

  \setlength{\tabcolsep}{3pt}
	\begin{tabular}{l|l||ccccc} 
         & $m$ & $2^{21}$ & $2^{22}$ & $2^{23}$ & $2^{24}$ & $2^{25}$  \\ \hline
        Time & GenCAT & $1.90e2$ & $3.77e2$ & $7.50e2$ & $1.56e3$ & $3.16e3$ \\  
    
        [sec] & LFR & $2.17e2$ & $7.66e2$ & $4.19e3$ & $1.50e4$ & $6.05e4$  \\ 
        & DC-SBM & $9.60e3$ &OOM&OOM&OOM&OOM\\ \hline
        Memory & GenCAT & $9.24e2$ & $1.78e3$ & $3.51e3$ & $7.33e3$ & $1.48e4$  \\ 
        
        [MiB] & LFR & $1.40e3$ & $2.25e3$ & $3.99e3$ & $7.45e3$ & $1.39e4$ \\
        & DC-SBM & $6.37e5$ & OOM & OOM & OOM & OOM \\ \hline \hline
 	\end{tabular}
 	  \setlength{\tabcolsep}{3pt}
	\begin{tabular}{l|l||cccccccccc} 
         & $m$ & $2^{26}$ & $2^{27}$ & $2^{28}$ & $2^{29}$ & $2^{30}$ \\ \hline
        Time & GenCAT & $6.53e3$ & $1.34e4$ & $2.84e4$ & $5.92e4$ & $1.23e5$ \\  
        
        [sec] & LFR & TO & TO & TO & TO & TO \\ 
        & DC-SBM &OOM&OOM&OOM&OOM&OOM\\ \hline
        Memory & GenCAT & $2.95e4$  & $5.92e4$ & $1.19e5$ & $2.37e5$ & $4.92e5$ \\ 
        
        [MiB] & LFR & TO & TO & TO & TO & TO \\
        & DC-SBM & OOM & OOM & OOM & OOM & OOM \\ 
 	\end{tabular} 	
 \end{center}
\end{table*}

To investigate the scalability of GenCAT, we demonstrate the runtime and memory consumption with varying the number of edges. 
We vary the number of edges $m$ within the range of 
$\{2^{21},2^{22},2^{23},2^{24},2^{25},2^{26},2^{27},2^{28},2^{29},2^{30}\}$.
We set the parameters $k$, $r$, and $n$ to $5$, $50$, and $m/32$, respectively, all diagonal elements of $\bmM$ to $0.6$, the other elements of $\bmM$ to $0.1$, all diagonal elements of $\bmD$ to $0.2$, and the other elements of $\bmD$ to $0.1$. 
We compare GenCAT with LFR and DC-SBM, which are the state-of-the-art generators closest in functionality to GenCAT.\footnote{We do not compare with TrillionG that is the state-of-the-art method in terms of scalability. The graphs generated by TrillionG are significantly different from those generated by GenCAT since TrillionG is a schema-driven approach without attributes and without flexible control of the class structure.} 

We first evaluate runtime of GenCAT, LFR, and DC-SBM for topology structure generation; recall that LFR and DC-SBM do not support attribute generation. 
In Table~\ref{tb: complexity}, we show the runtime to generate edges. 
This table indicates that GenCAT scales linearly to the number of edges and can generate graphs with billion edges in a reasonable time (i.e., the generation for $2^{30}$ edges finishes in $35$ hours).
Comparing GenCAT with LFR and DC-SBM, GenCAT significantly outperforms them.
This is because GenCAT efficiently generate edges due to the inverse transform sampling, but LFR often fails to generate edges as the graph size increases, and DC-SBM's complexity is $O(n^2)$. 
From the results, we confirm that GenCAT efficiently and scalably generates graphs with class labels. 

Next, we analyze memory consumption. 
Table~\ref{tb: complexity} shows the memory usage to generate graphs. 
It shows that GenCAT scales linearly to the number of edges. 
GenCAT and LFR have similar memory consumption. 
We observe that DC-SBM operates with memory usage proportional to the square of the number of nodes.

In summary, these experiments validate that both time and space complexities of GenCAT are linear to the number of edges, it significantly outperforms the state-of-the-art, and it generates edges with the controlled structure of large sizes not possible to practically generate with the state-of-the-art.

\subsection{Q4: How precisely does GenCAT reproduce real-world graphs?}
\label{ssec:Q4}
To evaluate reproducibility of real graphs with GenCAT, we show how precisely GenCAT reproduces the real-world graph by comparing with existing methods.
We use GenCAT, VGAE, NetGAN, LFR, and DC-SBM\footnote{LFR and DC-SBM do not have a function to reproduce given graphs. So, for LFR we manually set parameters appropriately to reproduce given graphs. As for DC-SBM, we compute node degrees by the same way as GenCAT and input them as parameters for DC-SBM.} to reproduce the CORA dataset \cite{MotlS15}, which is a typical graph dataset often used in studies of community structure, e.g., \cite{kipf2016variational,bojchevski2018netgan}. 
In this graph, the numbers of nodes, edges, and classes are $2810$, $7981$, and $7$, respectively. 

First, we clarify and explain the strong points and drawbacks of GenCAT and existing methods by visualizing the adjacency matrices of CORA dataset and generated graphs. 
Second, we quantitatively evaluate the class structure in reproduced graphs by the class features introduced in Section~\ref{ssec: basic statistics}. 
Third, we also evaluate the class structure by community-related statistics which are commonly used \cite{bojchevski2018netgan}. 
Then, we demonstrate that GenCAT can generate various size graphs with the same class structure as a given graph. 
\seijirv{Finally, we validate that GenCAT can reproduce the class structure of a large dataset. }

\begin{figure}[t]
  \centering
     \begin{minipage}{0.3\hsize}
      \centering
          \includegraphics[trim={0mm 0mm 0mm 0}, width=3.0cm]{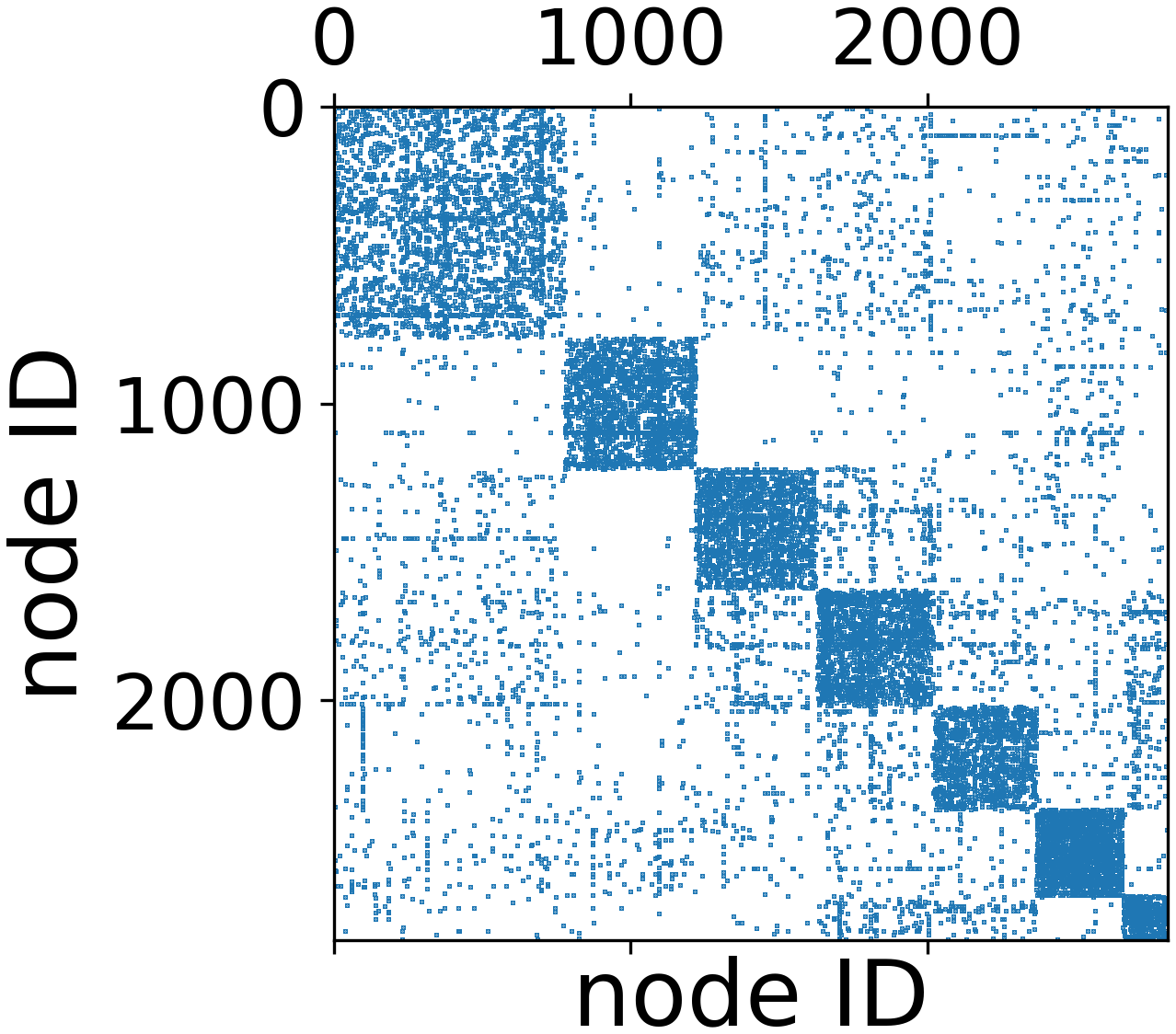}
          \subcaption{Original.}
          \label{fg: cora_original}
     \end{minipage}
     \begin{minipage}{0.3\hsize}
      \centering
          \includegraphics[trim={0mm 0mm 0mm 0}, width=3.0cm]{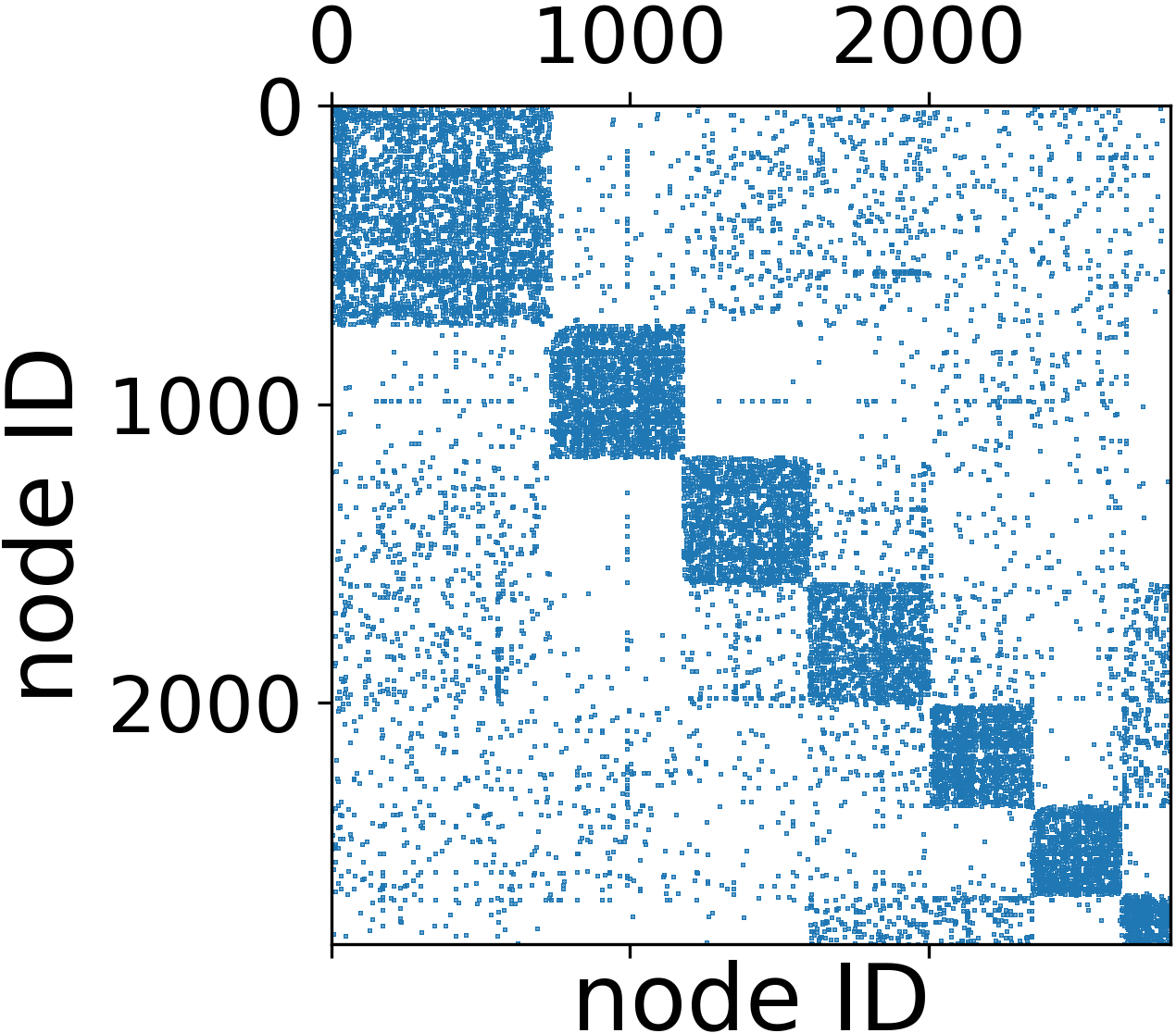}
          \subcaption{GenCAT.}
          \label{fg: cora_GenCAT}
     \end{minipage}
     \begin{minipage}{0.3\hsize}
      \centering
          \includegraphics[trim={0mm 0mm 0mm 0}, width=3.0cm]{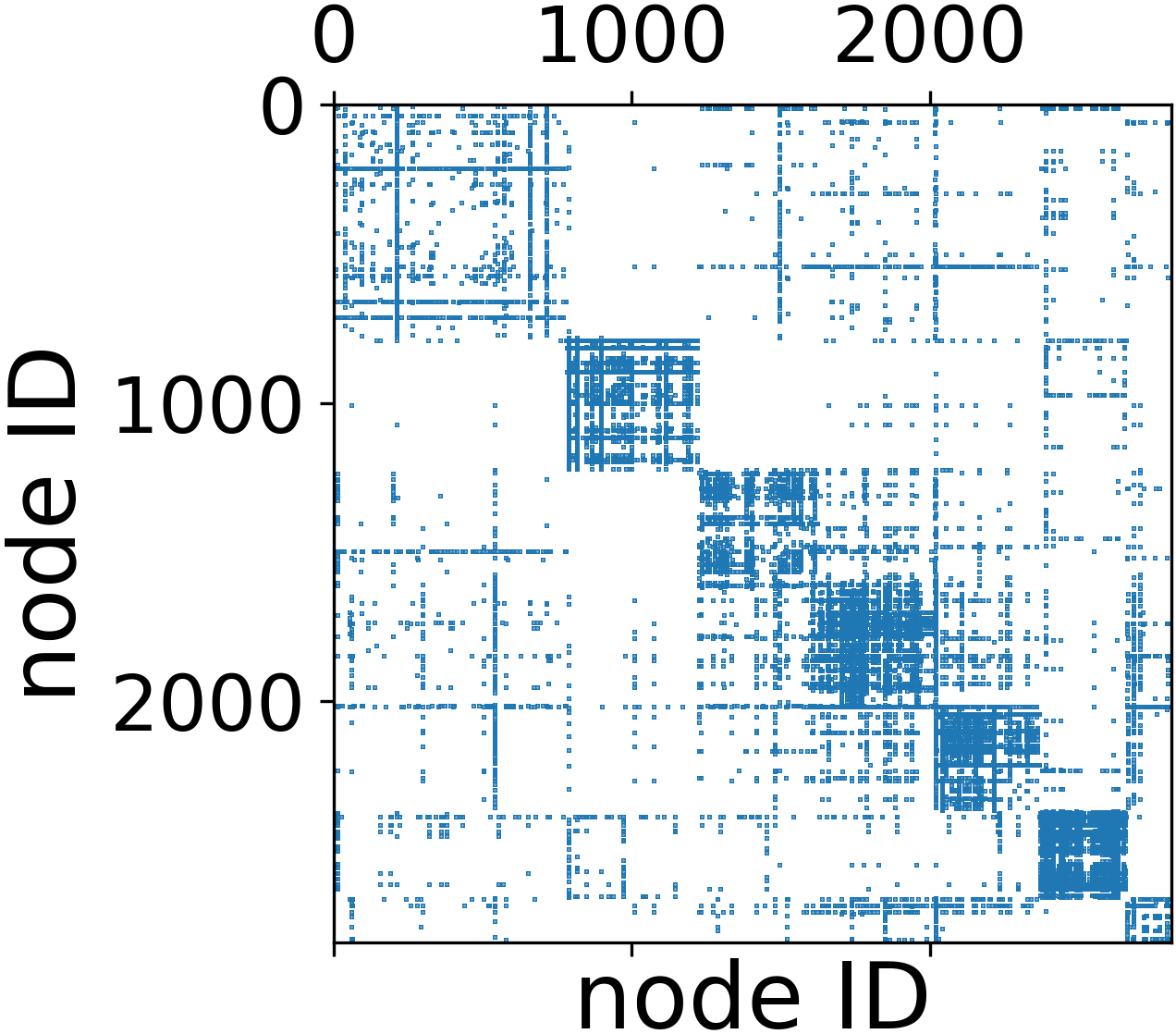}
          \subcaption{VGAE.}
          \label{fg: cora_gae}
     \end{minipage}
     \begin{minipage}{0.3\hsize}
      \centering
          \includegraphics[trim={0mm 0mm 0mm 0}, width=3.0cm]{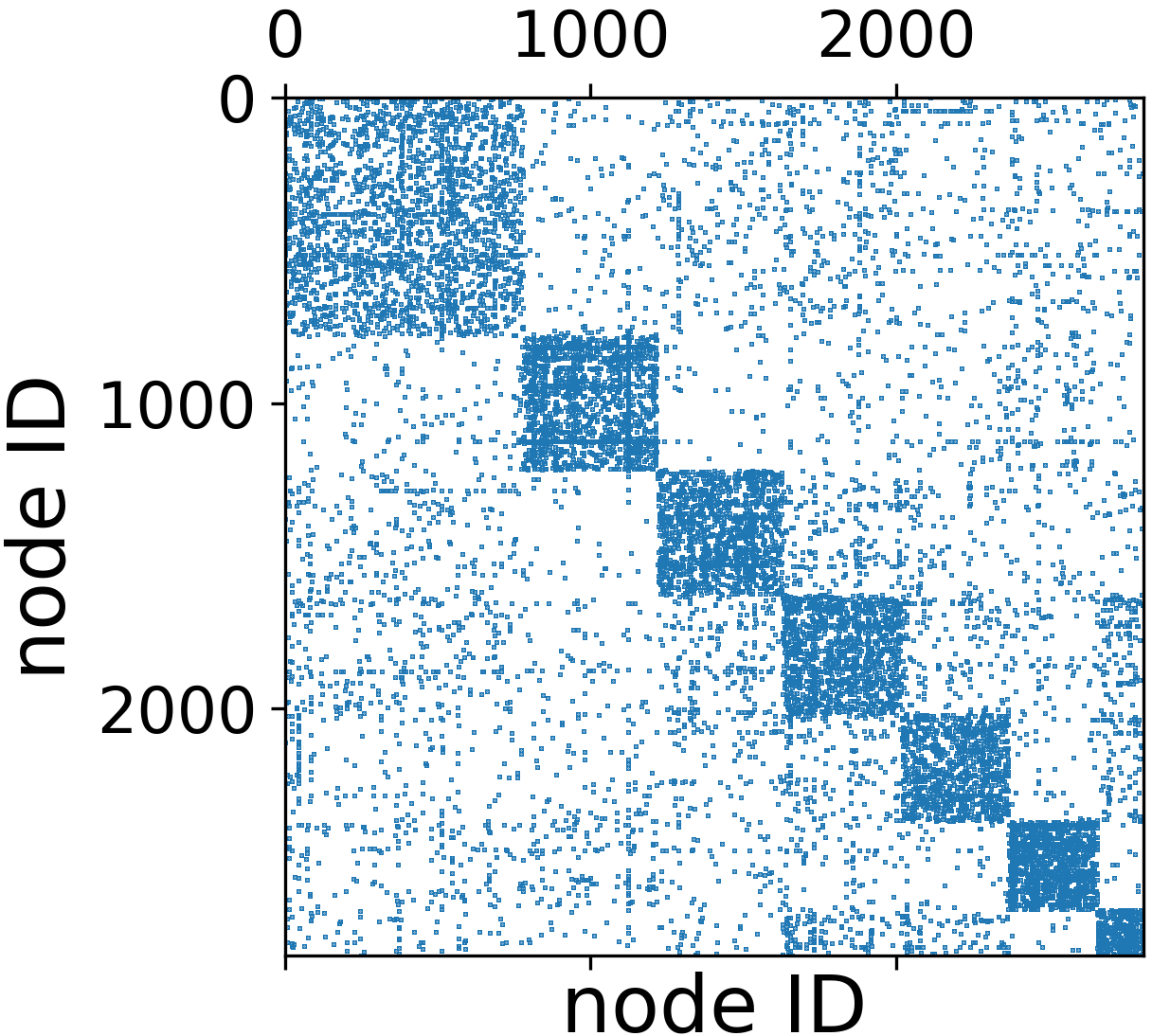}
          \subcaption{NetGAN.}
          \label{fg: cora_netgan}
     \end{minipage}
     \begin{minipage}{0.3\hsize}
      \centering
          \includegraphics[trim={0mm 0mm 0mm 0}, width=3.0cm]{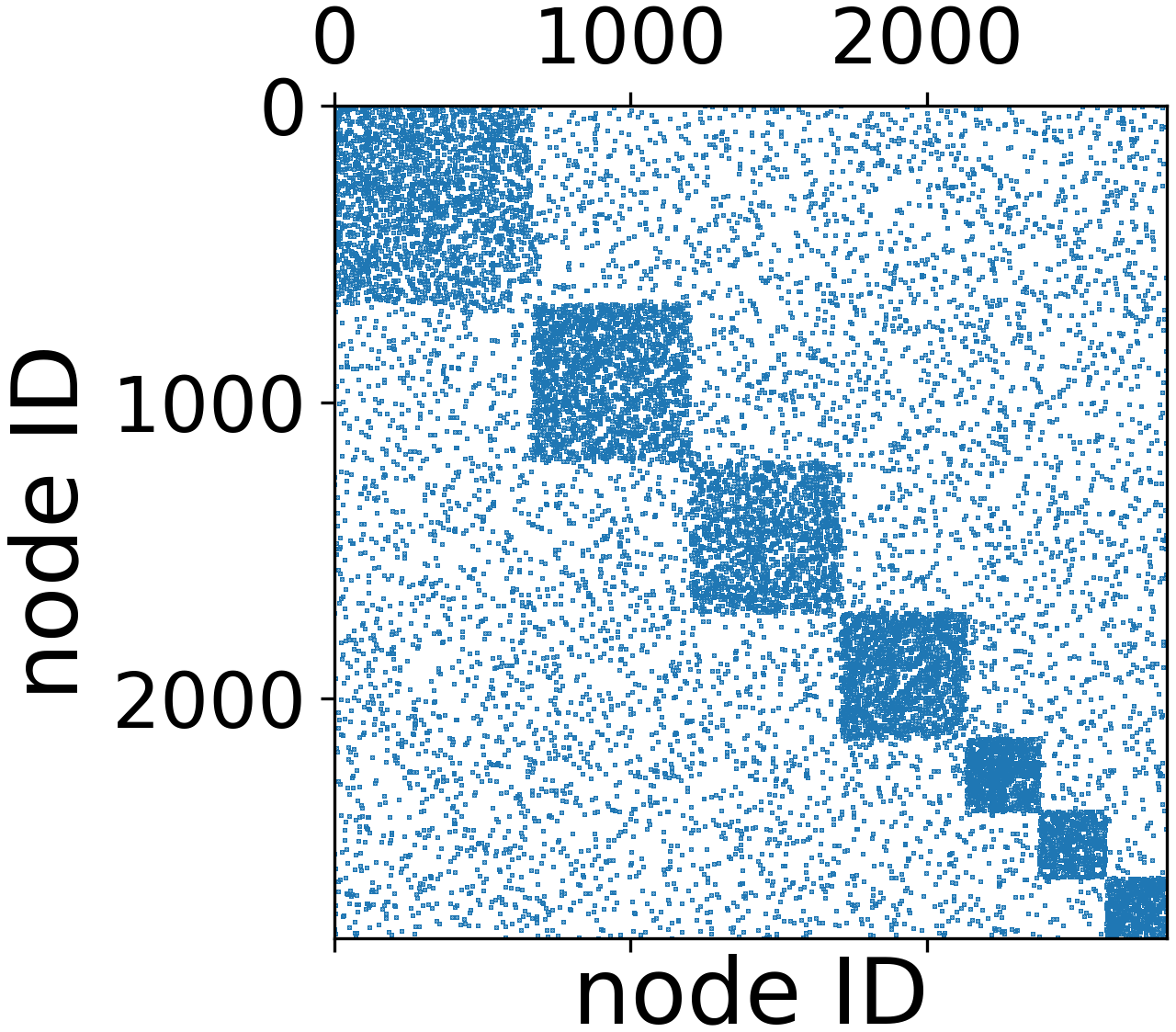}
          \subcaption{LFR.}
          \label{fg: cora_lfr}
     \end{minipage}
     \begin{minipage}{0.3\hsize}
      \centering
          \includegraphics[trim={0mm 0mm 0mm 0}, width=3.0cm]{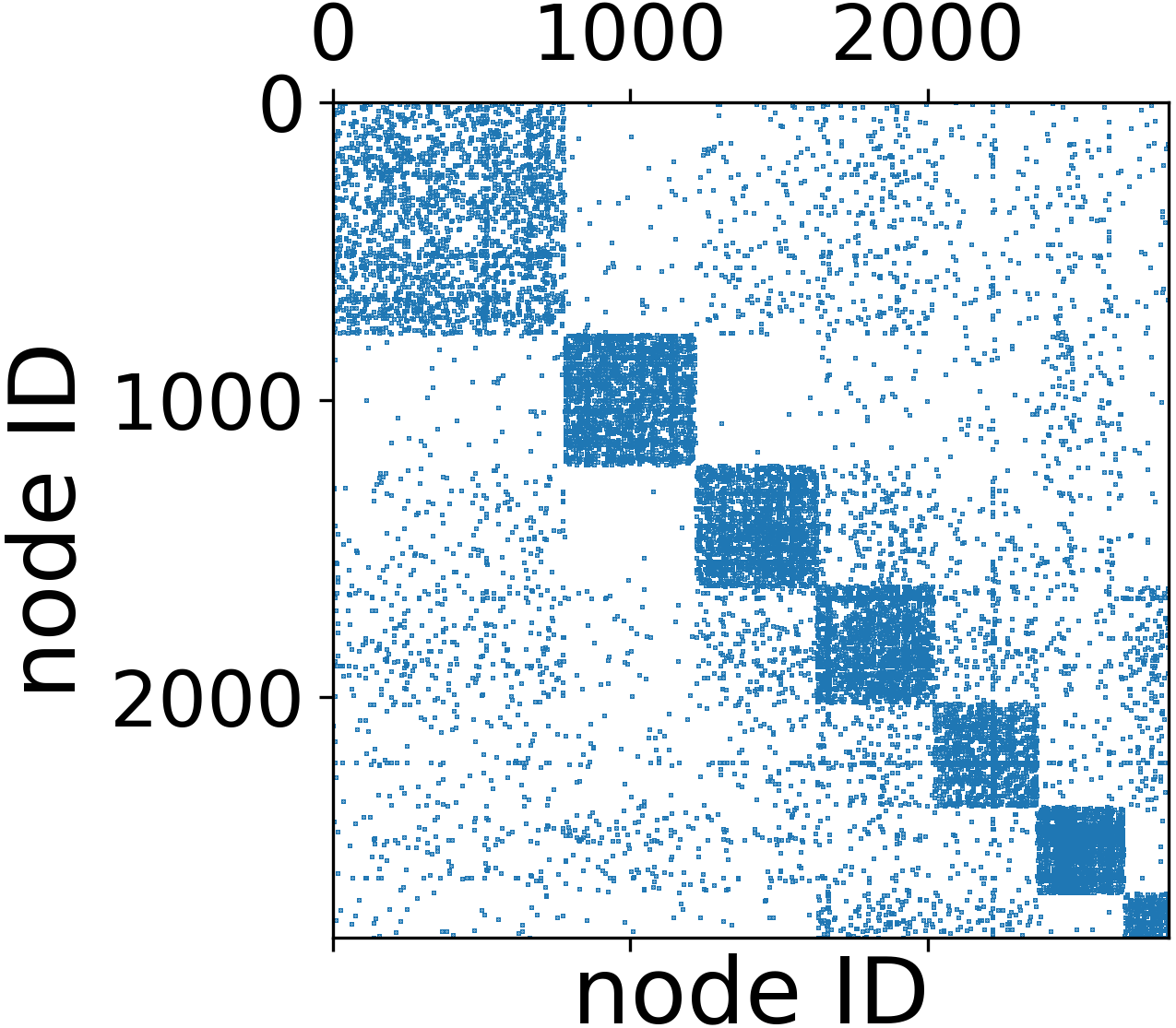}
          \subcaption{DC-SBM.}
          \label{fg: cora_sbm}
     \end{minipage}
    \caption{Visualization of adjacency matrices of CORA dataset and generated graphs.}
  \label{fg: reproduction}
\end{figure}
\subsubsection{Visualization}
Figure~\ref{fg: reproduction} shows the adjacency matrices of CORA dataset and generated graphs. 
For a fair comparison, we randomly order the nodes in each class. 
We here note that since VGAE and NetGAN do not explicitly generate the class labels and the nodes in generated graphs bijectively correspond to the nodes in original graphs, we assign nodes of the generated graph to the class labels of the corresponding nodes of the original dataset. 
From Figure \ref{fg: reproduction}, we observe that GenCAT most precisely reproduces CORA among the five methods.
This is because GenCAT can capture the connection proportions between nodes and classes. 
VGAE does not precisely reconstruct the graph structure since its main purpose is learning node embeddings. 
NetGAN does not explicitly consider the class structures in its learning steps, so it fails to reproduce the detailed parts. 
In LFR, inter-edges are uniformly distributed because LFR randomly generates them. 
In DC-SBM, inter-edges of each pair of classes are uniformly distributed since DC-SBM assumes that the nodes in the same class have the same connection proportions. 

\begin{small}
\begin{table}[t]
  \caption{Evaluation for class preference mean and class preference deviation between the original graph and generated graphs. Mean squared error (MSE) is used for the evaluation measure.}
  \label{tb: MSE cluster}
  \begin{center}
  \setlength{\tabcolsep}{3pt}
	\begin{tabular}{l|ccccc}
        Mean squared error & GenCAT & VGAE & NetGAN & LFR & DC-SBM \\ \hline
        Class preference mean & $8.71e$--$4$ & $2.05e$--$2$ & $2.93e$--$3$ & $7.05e$--$3$ & $1.66e$--$3$ \\ 
        Class preference deviation & $9.26e$--$4$ & $4.12e$--$3$ & $1.86e$--$3$ & $5.36e$--$3$ & $1.42e$--$3$ \\ 
 	\end{tabular}
 \end{center}
\end{table}
\end{small}
\subsubsection{Evaluation on class features}
To evaluate the accuracy quantitatively, we measure the mean square errors (MSE) of the class preference mean and class preference deviation between the original graph and generated graphs. 
Table~\ref{tb: MSE cluster} shows the MSE of all the methods. 
GenCAT achieves the best performance, and this result indicates that GenCAT can precisely reproduce the class structures in a given real graph. 

\begin{table}[!t]
  \caption{Statistics of Cora and the graphs generated by GenCAT and the baselines, averaged over five trials.}
  \label{tb:04_statistics}
  \begin{small}
    \begin{center}
    \setlength{\tabcolsep}{2.3pt}
      \begin{tabular}{l||c|ccccc} 
         & Original & GenCAT & VGAE & NetGAN & LFR & DC-SBM \\ \hline\hline
        Intra-community density & $ 1.97e$-$3$ & $ 1.89e$-$3$ & $ 2.04e$-$3$ & $ 1.41e$-$3$ & $ 1.36e$-$3$ & $ 1.88e$-$3$ \\
        Inter-community density & $ 5.07e$-$4$ & $ 4.92e$-$4$ & $ 4.87e$-$4$ & $ 6.0e$-$4$ & $ 7.03e$-$4$ & $ 5.01e$-$4$ \\
        Size of LCC & $ 2810.0 $ & $ 2810.0 $ & $ 1686.4 $ & $ 2804.2 $ & $ 2810.0 $ & $ 2499.0 $ \\
        \# connected components & $ 1.0$ & $ 1.0$ & $ 1123.4$ & $ 3.4$ & $ 1.0$ & $ 306.2$ \\
        Characteristic path length & $ 5.27$ & $ 4.36$ & $ 3.99$ & $ 5.19$ & $ 4.74$ & $ 4.14$ \\
         \hline
        Average rank & - & $ 1.8 $ & $ 3.8 $ & $ 3.0 $ & $ 2.8 $ & $ 3.2 $
      \end{tabular}
    \end{center}
  \end{small}
\end{table}
\subsubsection{Evaluation on community-related statistics}
We evaluate the quality of generated graphs by statistics related to communities, in order to show that GenCAT can generate realistic communities in a graph. 
The statistics include intra-community density, inter-community density, the size of largest connected component (called LCC for short), the number of connected components, and characteristic path length (average number of steps along the shortest paths for all node pairs), which are used in \cite{bojchevski2018netgan}. 
We observe that graphs generated by GenCAT are similar to the original graph in terms of all community-related statistics. 
The bottom row in Table \ref{tb:04_statistics} shows the average rank of each method over all statistics and demonstrates that GenCAT ranks the highest. 
This result means that GenCAT reproduces graphs with similar statistics to the original one. 
Though VGAE and DC-SBM precisely capture intra- and inter-community densities, they generate highly disconnected graphs. 
NetGAN achieves the closest score for characteristic path length because it learns random walks on a given graph. 
However, NetGAN cannot accurately capture intra- and inter-community density because it does not explicitly learn communities in a given graph. 
LFR fails to accurately reproduce intra- and inter-community density from the original graph since it assumes that all communities have the same density. 
In summary, we validate that GenCAT can generate realistic communities in a graph with regard to various statistics used commonly. 

\begin{figure}[t]
  \centering
     \begin{minipage}{0.48\hsize}
      \centering
          \includegraphics[trim={0mm 0mm 0mm 0}, width=4.cm]{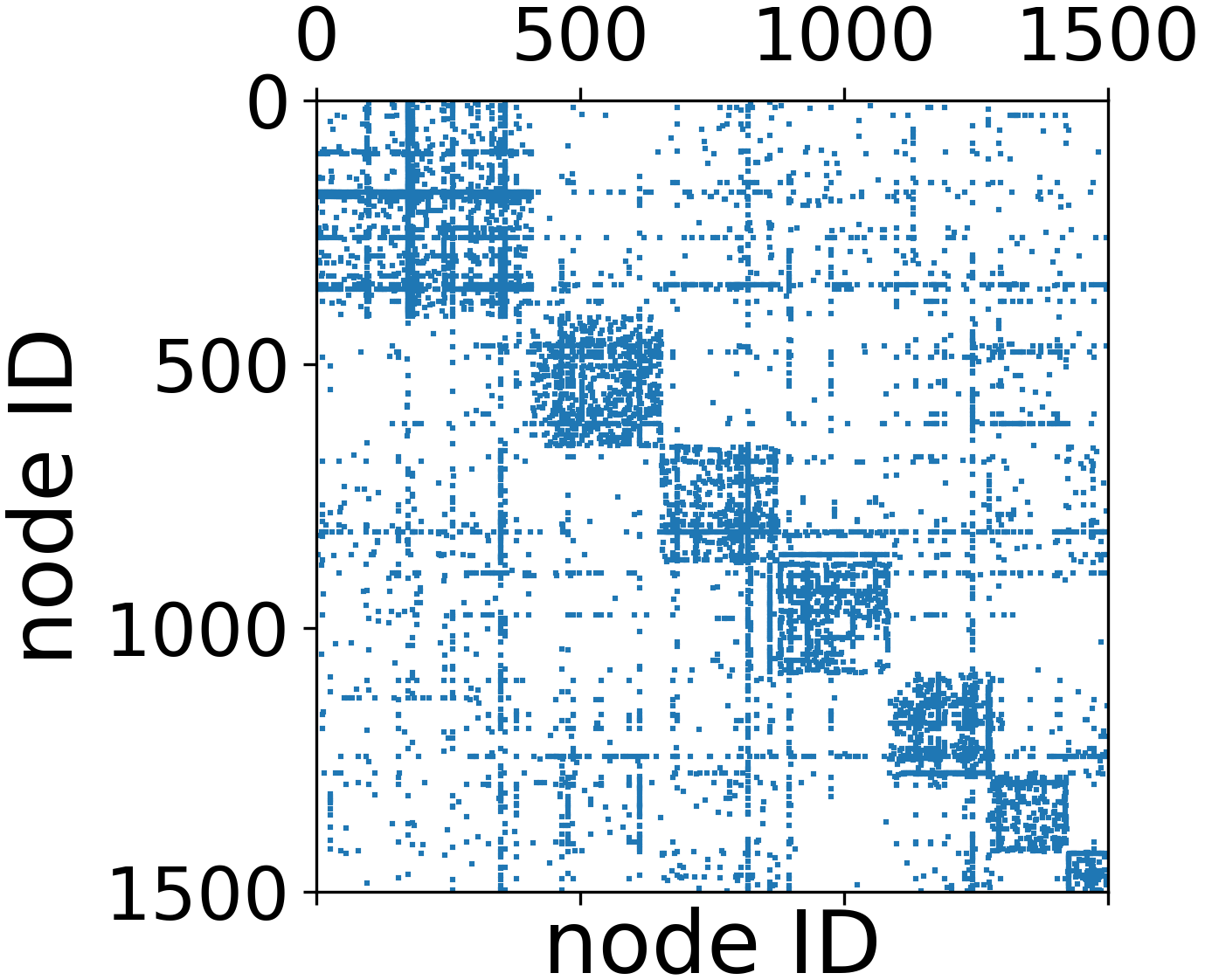}
          \subcaption{$n=1500, m=5000$.}
          \label{fg: small cora}
     \end{minipage}
     \begin{minipage}{0.48\hsize}
      \centering
          \includegraphics[trim={0mm 0mm 0mm 0}, width=4.cm]{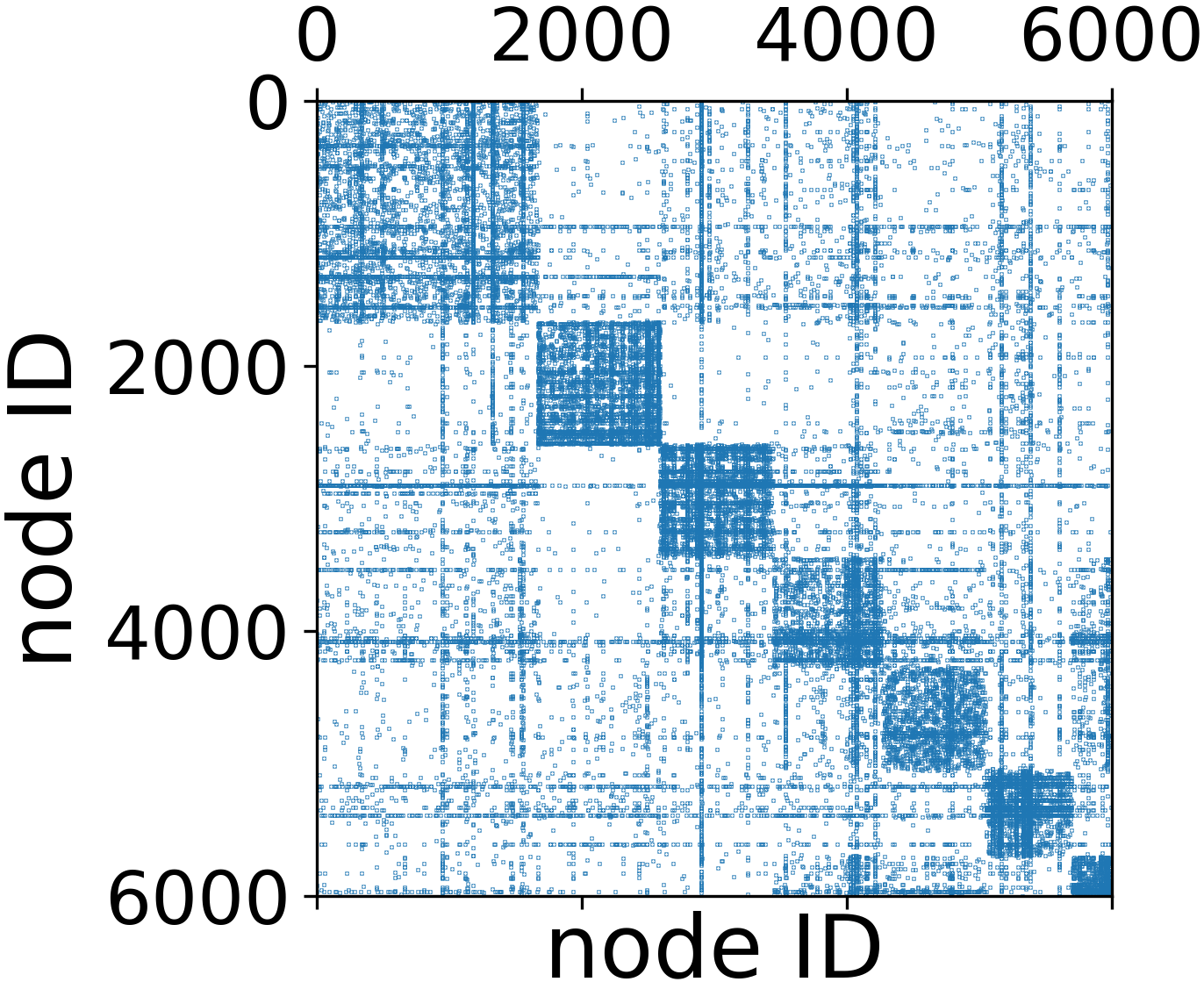}
          \subcaption{$n=6000,m=20000$.}
          \label{fg: large cora}
     \end{minipage}
    \caption{Visualization of adjacency matrices of generated graphs with various sizes.}
  \label{fg: chaging size}
\end{figure}
\subsubsection{Changing graph size}
Next, we demonstrate GenCAT generates various size graphs which have the same class structures as a given graph. 
Figure~\ref{fg: chaging size} shows two adjacency matrices of graphs generated by GenCAT: the parameters of the former are $n=1500$ and $m=5000$, and those of the latter are $n=6000$ and $m=20000$. 
Figure~\ref{fg: small cora} shows the reproduced graph has the similar class structure to CORA dataset shown in Figure~\ref{fg: cora_original}\footnote{We note that the adjacency matrix looks more sparse than the original dataset due to the size of the figure.}.
Figure~\ref{fg: large cora} shows that the larger graph also has a similar class structure to the original dataset.

These experiments validate that GenCAT can generate the class structures of real-world graphs and enable users to choose arbitrary sizes of generated graphs while maintaining the accuracy of these structures.

\begin{figure}[t]
  \centering
     \begin{minipage}{0.48\hsize}
      \centering
          \includegraphics[trim={0mm 0mm 0mm 0}, width=6.cm]{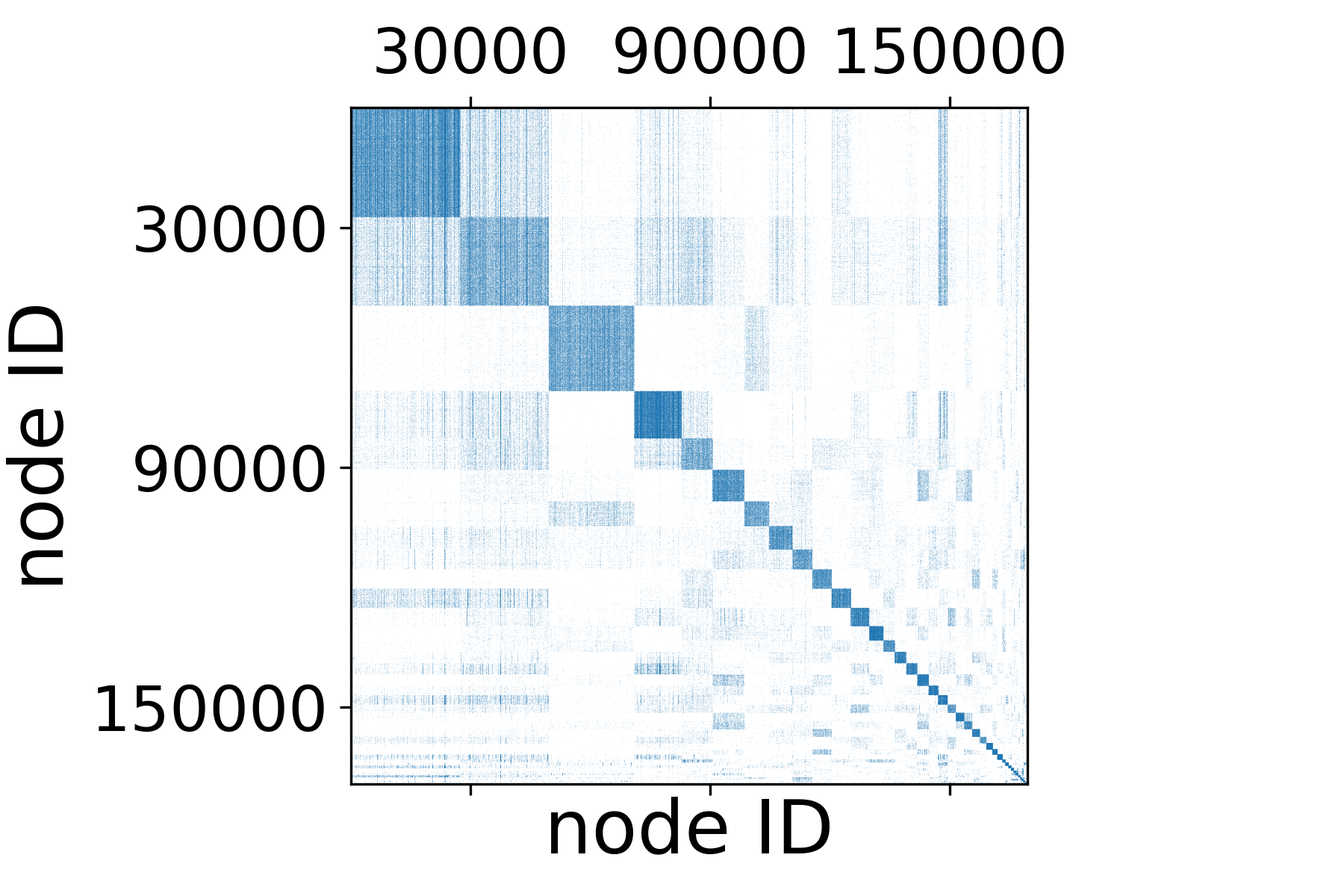}
          \subcaption{Original.}
          \label{fg: arxiv original}
     \end{minipage}
     \begin{minipage}{0.48\hsize}
      \centering
          \includegraphics[trim={0mm 0mm 0mm 0}, width=6.cm]{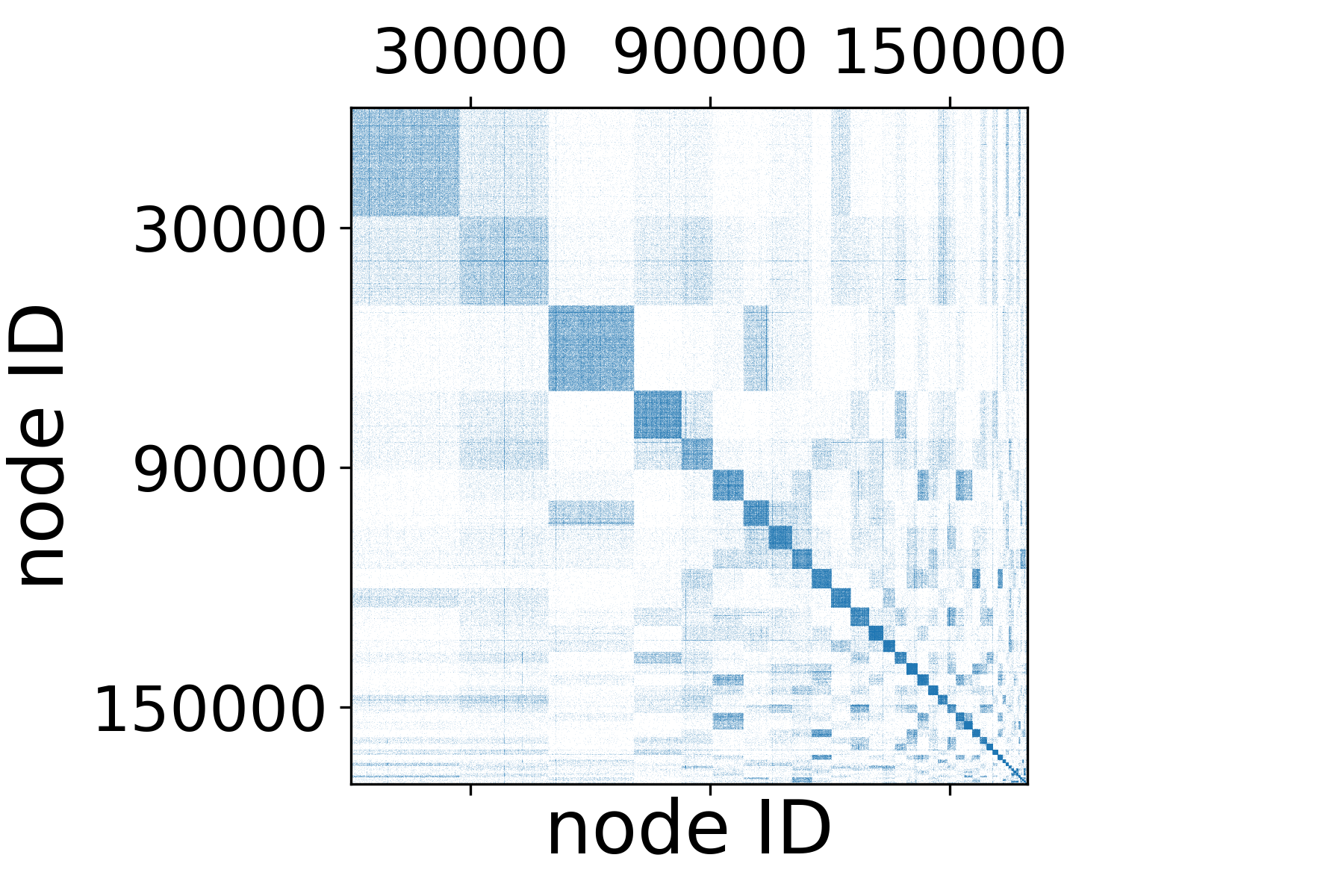}
          \subcaption{GenCAT.}
          \label{fg: arxiv gencat}
     \end{minipage}
    \caption{Visualization of adjacency matrices of ogbn-arxiv and a generated graph by GenCAT.}
  \label{fg: reproduction arxiv}
\end{figure}

\subsubsection{\seijirv{Reproduction of a large graph}}
\seijirv{Finally, in order to validate that GenCAT can reproduce the class structure of a large network, ogbn-arxiv~\cite{hu2020ogb}, which has 169,343 nodes, 1,166,243 edges, and 40 classes, we visualize the adjacency matrices of ogbn-arxiv dataset and a generated graph by GenCAT in Figure \ref{fg: reproduction arxiv}.
In this experiment, we use parameters extracted from ogbn-arxiv dataset for GenCAT's graph generation. Figure \ref{fg: arxiv original} shows that classes are densely connected internally in ogbn-arxiv. 
Figure \ref{fg: arxiv gencat} shows that GenCAT can generate graphs that are similar to the original. 
In summary, we observe that GenCAT can reproduce the class structure of a large dataset with over a million edges.}

\begin{table}[t]
  \caption{Ablation study on edge generation. We show execution time, MSE between user specified class preference mean and the class preference mean of a generated graph, and MSE between user specified class preference deviation and the class preference deviation of a generated graph. ``w/o ITS" and ``w/o AP" indicate the variants removing inverse transform sampling and adjusting proportion, respectively. We report the scores averaged over five trials. }
  \label{tb:04_ablation}
   \vspace{-3mm}
  \begin{center}
  \setlength{\tabcolsep}{3pt}
	\begin{tabular}{l|c|ccccc} 
        & $m$ & $2^{16}$ & $2^{17}$ & $2^{18}$ & $2^{19}$ & $2^{20}$ \\ \hline\hline
 & GenCAT & $ 6.34 $ & $ 12.30 $ & $ 23.78 $ & $ 47.71 $ & $ 95.29 $ \\
Time[sec] & w/o ITS & $ 30.70 $ & $ 110.10 $ & $ 421.84 $ & $ 1683.24 $ & $ 6714.91 $ \\
 & w/o AP & $ 6.01 $ & $ 11.52 $ & $ 22.15 $ & $ 44.87 $ & $ 89.87 $ \\ \hline
Class & GenCAT & $ 1.25e$-$3$ & $ 0.42e$-$3$ & $ 0.20e$-$3$ & $ 0.14e$-$3$ & $ 0.09e$-$3$ \\
preference & w/o ITS & $ 1.19e$-$3$ & $ 0.49e$-$3$ & $ 0.24e$-$3$ & $ 0.16e$-$3$ & $ 0.09e$-$3$ \\
mean & w/o AP & $ 14.96e$-$3$ & $ 13.57e$-$3$ & $ 13.32e$-$3$ & $ 12.96e$-$3$ & $ 12.93e$-$3$ \\ \hline
Class & GenCAT & $ 2.71e$-$3$ & $ 2.21e$-$3$ & $ 2.09e$-$3$ & $ 2.14e$-$3$ & $ 2.02e$-$3$ \\ 
preference & w/o ITS & $ 3.05e$-$3$ & $ 2.35e$-$3$ & $ 2.30e$-$3$ & $ 1.99e$-$3$ & $ 2.11e$-$3$ \\ 
deviation & w/o AP & $ 5.22e$-$3$ & $ 4.89e$-$3$ & $ 4.89e$-$3$ & $ 4.85e$-$3$ & $ 4.82e$-$3$ \\ 
 	\end{tabular}
 \end{center}
\end{table}

\subsection{Q5: How effective and efficient are proposed techniques on edge generation?} \label{ssec:Q5}
To validate the efficiency of inverse transform sampling (ITS) and the effectiveness of adjusting proportion (AP),
We evaluate the effect on edge generation of ITS and AP. 
We vary the number of edges $m$ within the range of 
$\{2^{16},2^{17},2^{18},2^{19},2^{20}\}$.
We set the parameters $k$, $r$, and $n$ as $6$, $50$, and $m/16$, respectively, all diagonal elements of $\bmM$ to $0.6$, and the other elements to $0.08$. 
Also, we set the diagonal elements of $\bmD$ to $[0.2,0.2,0.25,0.25,0.3,0.3]$, respectively, and the other elements to $0.05$. 

Table~\ref{tb:04_ablation} shows the execution time and losses between user specified class preference mean/deviation and the class preference mean/deviation of a generated graph. 
We use MSE in order to measure the losses between those class preference means/deviations. 
In this table, ``w/o ITS" and ``w/o AP" indicate the variants removing ITS and AP, respectively. 
First, the table shows that GenCAT and w/o AP scale linearly to the number of edges. 
On the other hand, w/o ITS requires $O(n^2)$ because it calculates the matrix multiplication $\bmU{\bmU'}^\top$ to obtain edge probability. 
Hence, we validate that ITS accelerates the edge generation of GenCAT. 
Second, in Table~\ref{tb:04_ablation} we observe that GenCAT and w/o ITS can generate graphs that satisfy the constraints of class preference mean/deviation better than w/o AP. 
From this observation, we validate that AP largely reduces the losses between user specified class preference mean/deviation and the class preference mean/deviation of a generated graph. 
Since the losses of GenCAT and w/o ITS are comparable, this result indicates that ITS does not decrease the quality of generated graphs.  

In summary, we conclude that ITS improves the efficiency of GenCAT while keeping the quality of edge generation in terms of class preference mean/deviation. 
Also we validate that AP improves the quality of edge generation.

\section{Related work}
\label{sec:related}
There is a rich literature on graph generation (e.g., \cite{chakrabarti2004r,leskovec2010kronecker,angles2014linked,bagan2017gmark,girvan2002community}). 
In this section, we first review \seiji{five} types of graph generators; \seiji{traditional generators,} generators for graphs with community structure, generators for graphs with community structure and node attributes, generators for large scale graphs, and neural network-based graph generators.
As described in Table \ref{tb: property}, we can see the advantages of GenCAT relative to the state of the art. 
\seiji{Then, we discuss the relationship between graph generators and a null model \cite{Katharina2016network} that has an unbiasedly random structure. }

\subsection{Existing graph generator}

\vspace{0.2cm}
\noindent\seiji{{\bf Traditional graph generator.}
While many traditional graph generators have been proposed such as Erd{\H{o}}s-R{\'e}nyi~\cite{erdHos1960evolution}, Barabasi-Albert~\cite{albert2002statistical}, Chung-Lu~\cite{aiello2000random}, and BTER~\cite{seshadhri2012community}, they cannot control the class structure in generated graphs. 
As for Erd{\H{o}}s-R{\'e}nyi, users can specify only edge density for a whole graph. 
Barabasi-Albert assumes that degree distributions follow power law distributions and implements a preferential attachment process so that generated graphs have power law node degree distributions. 
Chung-Lu aims to recreate a given node degree sequence. 
BTER controls degree distributions and cluster coefficient in generated graphs. 
In summary, these graph generators explicitly control edge density or node degree distributions but ignore the class structure in generated graphs. 
}

\vspace{0.2cm}
\noindent{\bf Generator for graphs with community structure.}
This type of graph generators takes into account not only the topological characteristic of a complex graph (e.g., power law node degree) but also topological structures within communities.
The LFR-benchmark~\cite{lancichinetti2008benchmark} is designed to evaluate community mining algorithms. 
This assumes that the distributions of node degrees and community sizes follow power law distributions. 
The LFR-benchmark is extended to generate synthetic graphs with overlapping communities (i.e., nodes belong to multiple communities)~\cite{sengupta2017benchmark} and hierarchical communities \cite{yang2017hierarchical}. 
DC-SBM~\cite{holland1983stochastic,karrer2011stochastic} supports controlling the proportions of connections between classes. 
However, since this type of graph generators does not support node attributes, it cannot capture relationships between attributes and topological structures. 
In addition, these generators only consider the class-level connection proportions, so nodes in a class have the same node-class membership proportions. 

\vspace{0.2cm}

\noindent{\bf Generator for graphs with community structure and node attributes.}
There are few generators that take both community structure and node attributes into account.
This type of graph generators generates edges between nodes according to the similarity of their attributes.
ANC~\cite{largeron2015generating} is a generator for attributed graphs with community structure, and DANCer~\cite{benyahia2016dancer} is an extended generator of ANC for generating dynamic graphs.
In the design of ANC, the community structure of generated graphs only depends on the attributes.
This indicates that it cannot flexibly generate a variety of graphs because users cannot explicitly control the connection proportions for each class.
In GenCAT, users can flexibly control the connection proportions between nodes and classes in graphs. 

\vspace{0.2cm}
\noindent{\bf Generators for large scale graphs.}
There are a number of generators for large scale graphs~\cite{chakrabarti2004r,leskovec2010kronecker,bagan2017gmark,park2017trilliong}. 
gMark~\cite{BBCFLA16gmark,bagan2017gmark}, pgMark~\cite{pgmark2018}, and TrillionG~\cite{park2017trilliong} are schema-driven graph generation methods that support node labels and edge predicates. 
pgMark allows users to flexibly generate graphs by leveraging an optional schema definition, called a graph configuration, and supports node attributes. 
TrillionG can generate large scale graphs efficiently by leveraging a recursive vector model. 
However, the generators cannot explicitly take community structure into account. 

\vspace{0.2cm}
\noindent{\bf Neural network-based graph generators.}
Recently, neural network-based graph generators~\cite{kipf2016variational,simonovsky2018graphvae,bojchevski2018netgan,you2018graphrnn,yang2019conditional,shi2020effective} have been developed to reproduce real-world graphs.
VGAE~\cite{kipf2016variational} and GraphVAE~\cite{simonovsky2018graphvae} construct generative models by leveraging a variational autoencoder. 
The main idea of them is that they consist of a graph encoder of GCN and a decoder that outputs an adjacency matrix. 
Also, NetGAN~\cite{bojchevski2018netgan} is proposed to learn the generation of walks from biased random walks instead of graphs. 
However, the existing graph generators aim to reproduce synthetic graphs from given input graphs, so these generators cannot flexibly generate various graphs because users cannot explicitly control the characteristics of generated graphs. 
Additionally, they cannot generate large scale graphs due to the large training time of their models.

\seiji{
\subsection{The use of null models in the study of graph generators}
A typical way to analyze empirical data is to compare it with a randomized version of the data, often called \textit{null model} \cite{Katharina2016network}. 
As for a graph generation problem, given the same numbers of nodes and edges as the original data, the edges in a null model are generated unbiasedly in terms of node degrees and classes, i.e., all nodes have similar degrees and the same connection proportions for all classes. 
As we showed in our experiments, GenCAT can flexibly control the node degrees and class structure in generated graphs and the generated graphs can be biased by the node degrees and class structure. 
In this sense, graphs generated by GenCAT are different from the null model. 
Moreover, GenCAT can mimic a null model by appropriately setting its parameter, i.e., we set node degrees to the same value for all nodes and set class preference means to the same value for all pairs of classes. 
Graphs generated by other existing graph generators also differ from the null model in their focus. 
For example, LFR controls the ratio of intra- and inter-edges so the connection proportions between classes are biased in its generated graphs, unlike a null model. 
As for the traditional graph generators, they can generate graphs biased by user-specified node degree, and/or cluster coefficient. 
In summary, the use of a null model enables us to understand how generated graphs are biased compared with their randomized versions. 
}

\section{Concluding remarks}
\label{sec:conclusion}
We presented GenCAT, a generator for attributed graphs with class labels. 
We experimentally validated four major aspects of our generator: 
1) edges in generated graphs follow user-specified graph features (i.e., node degrees) and user-specified class features (i.e., the proportions of connections between nodes and classes), 
2) GenCAT can generate attributes that follow user-specified attribute distributions and the controlled class structure, 
3) GenCAT scales linearly to the number of edges, and we show that GenCAT generates graphs with billion edges, 
4) GenCAT more precisely reproduces the class structure in real-world graphs than existing methods. 
GenCAT is the first graph generation method having all four of these practical features.
Through our experiments, we demonstrated that GenCAT can successfully generate massive attributed graphs with sophisticated user-controlled class structures. 

Interesting directions for future work include extending GenCAT to generate directed graphs, overlapping communities, and hierarchical communities.
Also, it is an interesting topic to consider the graph generation problem as an inverse of the graph clustering problem. Here the goal would be to design a single unifying framework that explains both attributed graph generation and attributed graph clustering.

\section*{Acknowledgment}
This work was supported by JSPS KAKENHI Grant Numbers JP20H00583.

\bibliography{references}%

\end{document}